\documentclass[sigconf]{acmart}
\usepackage{xspace}
\usepackage{xcolor}
\usepackage{paralist}
\usepackage{wrapfig}
\usepackage{multirow}
\usepackage{listings}
\usepackage{makecell}
\usepackage{boxedminipage}
\usepackage{booktabs}
\usepackage{subcaption}
\usepackage{tabularx}
\usepackage{url}
\usepackage{amsmath}

\usepackage{breakurl} 
\usepackage{boldline}
\usepackage[ruled,vlined]{algorithm2e}
\usepackage[titletoc]{appendix}
\usepackage{pifont}\usepackage{balance}
\usepackage[htt]{hyphenat}
\usepackage{mdframed}

\keywords{Virtual Reality; Firmware Analysis}

\ccsdesc[500]{Security and privacy~Mobile platform security}
\ccsdesc[500]{Security and privacy~Software security engineering}

\copyrightyear{2025} 
\acmYear{2025} 
\setcopyright{acmlicensed}
\acmConference[CCS '25]{Proceedings of the 2025 ACM SIGSAC Conference on Computer and Communications Security}{October 13--17, 2025}{Taipei, Taiwan}
\acmBooktitle{Proceedings of the 2025 ACM SIGSAC Conference on Computer and Communications Security (CCS '25), October 13--17, 2025, Taipei, Taiwan}

\title[A Longitudinal Security Analysis of VR Firmware]{\textit{Virtual} Reality, \textit{Real} Problems: \\A Longitudinal Security Analysis of VR Firmware}

\author{Vamsi Shankar Simhadri}
\affiliation{\institution{Department of Computer Science\\George Mason University}
  \city{Fairfax, VA}
  \country{United States}
}
\email{vsimhadr@gmu.edu}

\author{Yichang Xiong}
\affiliation{\institution{Department of Computer Science \\George Mason University}
  \city{Fairfax, VA}
  \country{United States}
}
\email{yxiong2@gmu.edu}

\author{Habiba Farrukh}
\affiliation{\institution{Department of Computer Science\\ University of California, Irvine}
  \city{Irvine, CA}
  \country{United States}
}
\email{habibaf@uci.edu}

\author{Xiaokuan Zhang}
\affiliation{\institution{Department of Computer Science \\George Mason University}
  \city{Fairfax, VA}
  \country{United States}
}
\email{xiaokuan@gmu.edu}

\newcommand{\bheading}[1]{{\vspace{2pt}\noindent{\textbf{#1}}}}
 
\usepackage{comment}
\usepackage{array, tabularx}

\def\Snospace~{\S{}}

\newcommand{\cc}[1]{\mbox{\smaller[0.5]\texttt{#1}}}

\newcounter{note}[section]

\newcommand{\rev}[1]{\textcolor{black}{#1}}

\newcommand{\eg}{\emph{e.g.}\xspace}

\newcommand{\ignore}[1]{}

\newcounter{packednmbr}

\newenvironment{packeditemize}{
\begin{list}{$\bullet$}{
\setlength{\labelwidth}{6pt}
\setlength{\itemsep}{2pt}
\setlength{\leftmargin}{\labelwidth}
\addtolength{\leftmargin}{\labelsep}
\setlength{\parindent}{1pt}
\setlength{\listparindent}{\parindent}
\setlength{\parsep}{1pt}
\setlength{\topsep}{1pt}}}{\end{list}}
 
\begin{document}

\begin{abstract}
Virtual Reality (VR) technology is rapidly growing in recent years. 
VR devices such as Meta Quest 3 utilize numerous sensors to collect users' data to provide an immersive experience.
Due to the extensive data collection and the immersive nature, 
the security of VR devices is paramount.
Leading VR devices often adopt and customize Android systems,
which makes them susceptible to both Android-based vulnerabilities and new issues introduced by VR-specific customizations (e.g., system services to support continuous head and hand tracking).
While prior work has extensively examined the security properties of the Android software stack, how these security properties hold for VR systems remains unexplored.

In this paper, we present the {\it first} comprehensive security analysis of VR firmware. 
We collect over 300 versions of VR firmware from two major vendors, Quest and Pico,
and perform a longitudinal analysis across the kernel layer, the system binary and library layer, and the application layer.
We have identified several security issues in these VR firmware, including missing kernel-level security features,
insufficient binary hardening, inconsistent permission enforcement, and inadequate SELinux policy enforcement. Based on
our findings, we synthesize recommendations for VR vendors to
improve security and trust for VR devices.
This paper will act as an important security resource for VR developers, users, and vendors, and will also direct future advancements in secure VR ecosystem.

\end{abstract}
\maketitle

\section{Introduction}
\label{sec:intro}

\noindent
Virtual Reality (VR) technology is a rapidly growing trend with a projected market value of \$22 billion in 2025~\cite{vr-market}. One of the industry's leaders, Meta (formerly known as Facebook), has sold approximately 20 million VR headsets as of March 2023~\cite{quest-sale1,quest-sale2} and invested billions in its VR venture. Apple recently launched Vision Pro~\cite{visionpro}, 
a new mixed reality device,
and Google is expected to release the AndroidXR operating system in 2025~\cite{androidXR}.

Despite its widespread adoption, VR technology introduces critical security and privacy challenges. The extensive collection and processing of personal data by VR devices, including biometric identifiers, behavioral patterns, and environmental information, raises significant privacy concerns~\cite{vr-privacy-1,vr-privacy-2,vr-privacy-4}. The compromise of such sensitive data could lead to severe consequences, including identity theft, unauthorized system access, and invasion of personal privacy. Beyond data protection, VR systems present unique safety challenges~\cite{vr-safety-1,vr-safety-2} due to their immersive nature, which completely obscures users' perception of their physical surroundings~\cite{cheng2023exploring}.
To mitigate potential physical harm, VR devices implement runtime safeguards, such as virtual boundaries~\cite{guardian}, to prevent accidents like dangerous falls and collisions.
Yet, these safety mechanisms are mediated by the VR device's operating system and could be rendered ineffective if the software stack is compromised, leading to far-reaching implications for end-user well-being.

\looseness=-1

These risks underscore the importance of robust security in the software that powers VR platforms.
Leading VR platforms such as Meta Quest and Pico
have strategically adopted customized Android-based operating systems, leveraging Android's robust open-source architecture as a foundation.
These manufacturers undertake sophisticated customization processes to seamlessly integrate proprietary system software with the base operating system, resulting in device-specific {\bf firmware} that powers their cutting-edge products. While the core Android framework maintains its open-source nature, the customized firmware implementations remain proprietary and closed-source, creating significant barriers to transparency and independent security assessment.

Operating as critical downstream components in the {software supply chain}, these modified firmware systems face a dual security challenge: they inherit Android's inherent vulnerabilities while introducing novel security considerations through extensive customization. The Android Security Bulletin~\cite{android-security-bulletin} consistently reveals numerous critical security vulnerabilities requiring immediate patching, yet downstream device manufacturers face substantial challenges in implementing these crucial updates in a timely manner~\cite{zhang2018precise,zhang2021investigation}. This concern is further amplified by extensive research demonstrating that vendor-specific Android customizations frequently introduce additional security vulnerabilities~\cite{wu2013impact,zhou2014peril,aafer2016harvesting,yu2021sepal,possemato2021trust,el2021dissecting,liu2022customized,jin2023dependency}.
These risks are significantly amplified in VR systems, where firmware must support continuous, high-bandwidth interaction with complex sensors like eye trackers, and depth sensors as well as scene and spatial understanding pipelines.
Supporting such capabilities often necessitates architectural changes to the Android-based firmware, which may introduce novel vulnerabilities.

While prior security research on VR systems~\cite{miller2022combining,liebers2021understanding,olade2020biomove,pfeuffer2019behavioural,tricomi2022you,kumar2022passwalk,zhang2023s,nair2022exploring,slocum2023going, luo2022holologger,al2021vr,meteriz2022keylogging,su_remote_2024,luo_eavesdropping_2024,nguyen_penetration_2024,jarin2023behavr}
has primarily focused on application-level threats and user privacy concerns, such deanonymization via sensor data analysis, behavior profiling, and keystroke inference through side-channel attacks, the security of underlying VR firmware powering these systems remains largely unexplored.
As VR technology rapidly transitions from an emerging technology to mainstream adoption, there is a critical need to systematically investigate potential security vulnerabilities in VR firmware. 
This research gap is particularly concerning given VR's expanding role in sensitive domains like military training, healthcare, and enterprise environments, where firmware vulnerabilities could potentially compromise not only individual user safety but also organizational security.

In this paper, we seek to address this research gap by analyzing how VR vendors adopt security measures to protect VR firmware.
Android has evolved a rich set of protections, including kernel-level mitigations, binary hardening, access control policies and SELinux-based isolation~\cite{maar2024defects, smalley2013security, androidPermissions, android_native_code_risks}
to protect against various vulnerabilities.
However, it is unclear whether these defenses are preserved, degraded, or bypassed in VR firmware, where performance demands and proprietary customizations are intense.

\bheading{Challenges.}
Analyzing the firmware security for VR platforms, however, presents several challenges.
First, unlike Android smartphones, where root access is readily available, VR devices are designed to be tamper-resistant and do not support root access.
Second, firmware security-related configurations, such as kernel configurations, SEPolicies, or permission settings, are spread across multiple image layers and file paths in VR firmware.
Lastly, VR platforms lack standardized firmware structure.
For instance, Meta uses \cc{payload.bin}, while Pico uses Brotli-compressed partitions; some devices consolidate firmware customizations in \cc{system.img}, others split them across \cc{system.img}, \cc{vendor.img}, and \cc{odm.img}. This variability creates a fragmented firmware landscape where firmware analysis tools must be adapted for each target.

To address these challenges and fill the current research gap, we present the {\it first} comprehensive, longitudinal investigation into VR firmware security.
Our study employs a systematic approach to analyze VR firmware security across multiple architectural layers (kernel layer, binary layer, and application layer) through rigorous differential analysis against the Android Open Source Project (AOSP) baseline in relevant cases.
We have assembled an extensive dataset comprising over 300  firmware versions from two leading VR manufacturers, Meta and Pico, enabling us to conduct a thorough longitudinal analysis.
From a high-level, we unzip each firmware, and extract  image files that are important for our study; then,
we mount those images locally to extract necessary files (e.g., kernel binaries, preinstalled apps), and perform analysis.

\bheading{Findings.}
Our analysis of VR firmware reveals systemic security weaknesses across multiple layers of the software stack. 
At the kernel level, VR devices frequently suffer from misconfigurations and delayed adoption of up-to-date Linux versions, leading to privilege escalation attacks.
Similarly, binary hardening is inconsistently applied, with key protections like Control Flow Integrity (CFI) and Fortify Source often absent.
At the application level, improper security flag usage and flawed permission handling expose VR devices to risks such as unauthorized data backups, man-in-the-middle (MITM) attacks, and permission leaks.
Moreover, vendors frequently modify SEPolicies and leave system properties exposed to untrusted apps.
Collectively, these findings suggest that despite a growing market presence, VR devices lag behind in adopting standard mobile security practices.

\bheading{Contributions.} 
This paper makes the following contributions:
\begin{packeditemize}
    \item We present the first \rev{longitudinal, multi-layer investigation into VR firmware security. Our work complements prior studies on vendor customizations and firmware vulnerabilities for Android and IoT devices, but we focus on VR, where distinct risks stem from continuous real-time sensor input, immersive runtime dependencies, and proprietary firmware design.}
\item We develop a set of tools to extract and analyze VR firmware across multiple vendors and hardware versions, \rev{which are designed for fragmented and vendor-specific VR firmware formats.} The tools will be open-sourced to facilitate future research.
    \item We present several important findings regarding the security of VR firmware, including missing kernel-level security features, insufficient binary hardening, inconsistent permission enforcement, and inadequate SELinux policy enforcement. Based on our findings, we synthesize recommendations for VR vendors to improve security and trust for VR devices.
\end{packeditemize}

\bheading{Ethics consideration.}
The firmware dataset used in this paper was collected from public resources.
We have disclosed our findings with Meta and Pico teams, and their responses are discussed in~\autoref{sec:discuss}.

\bheading{Artifacts.}
Our code and data are available at:
\url{https://github.com/SECSAT-LAB-GMU/VR-Firmware}.

 \section{Background}
\label{sec:bg}
\subsection{VR Device Types}
Virtual Reality (VR) offers immersive 3D experiences via advanced head-mounted displays (HMDs) \cite{slater2016enhancing}. The combination of market opportunities and technology innovation has driven significant investments by major tech companies in proprietary VR platforms \cite{statistaVRMarket}. 
VR systems are categorized into four types based on architecture: 
(1) Basic Optical VR \cite{googleCardboardSpecs} (e.g., Google Cardboard), using passive optical elements like precision convex lenses; (2) Smartphone-tethered VR \cite{gearVRoverview} (e.g., Samsung Gear VR), using mobile devices with tracking sensors;
(3) PC-tethered VR \cite{indexTechSpecs} (e.g., Valve Index), utilizing high-performance PCs for graphics and physics; and (4) Standalone VR \cite{quest3Whitepaper} (e.g., Meta Quest 3), combining processing units, displays, and sensors into a self-contained platform. 
This paper focuses on standalone VR devices as the leading market segment and complete computational systems with comprehensive firmware. We also only focus on Android-based devices such as Meta Quest.

\begin{figure}[t]
    \centering
    \includegraphics[width=.9\columnwidth]{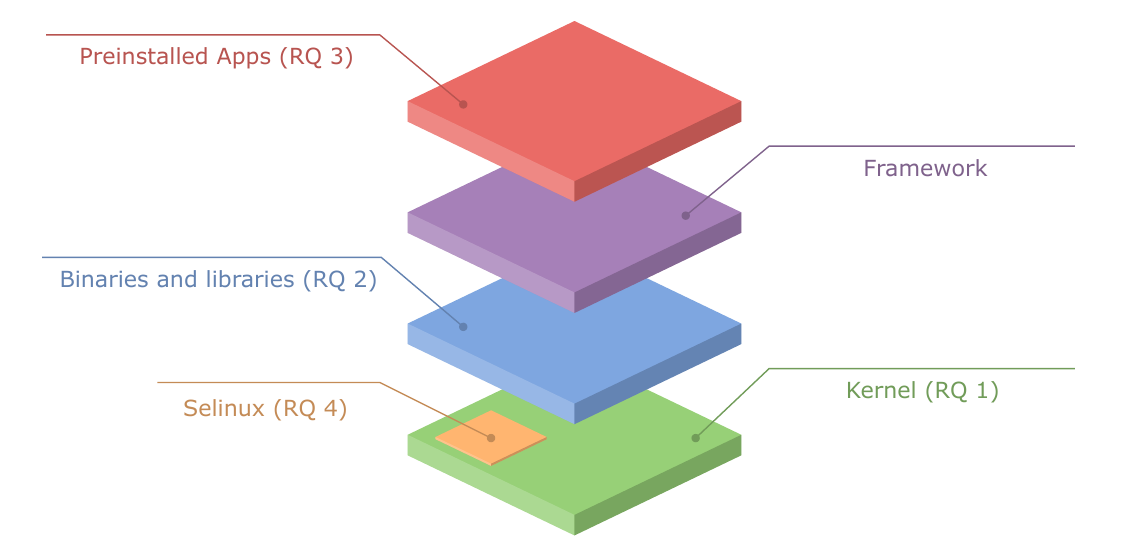}
\caption{VR Firmware}
    \label{fig:structure}
\end{figure}

\subsection{VR Firmware}

Most VR devices run customized Android firmware containing essential components, policies, and configurations for proper functioning. This firmware includes executables, libraries, pre-installed apps, and scripts for system setup.
All use at least Android 10, except Oculus Quest, which started with Android 7.1.1 and later upgraded to Android 10. 
Manufacturers enhance firmware with VR-specific binaries, services, and applications, adjusting functionalities. Customizations include integrating kernel components like VR drivers and refining security and performance for immersive experiences.
At a high level, VR firmware follows the following structure~\cite{android_arch_overview} (shown in~\autoref{fig:structure}), with vendors frequently modifying these components to meet their specific needs.

\bheading{Kernel Layer.}
The kernel  serves as the foundation of the VR device, providing essential system functionalities. The VR kernel layer consists of the core Android kernel with VR-specific modifications. These include VR-specific drivers, such as display drivers, GPU drivers, and sensor drivers, which ensure performance and responsiveness. Within the firmware, the kernel binary is stored in \cc{boot.img}.

\bheading{System Binaries and Libraries Layer.}
Binaries and libraries within the firmware are essential system components that provide pre-compiled functions and services required by both the operating system and applications. These components are crucial for system stability and performance. However, they are also a common source of critical security vulnerabilities~\cite{possemato2021trust}, 
as many are developed using unsafe programming languages like C and C++, which can introduce memory-related vulnerabilities.

\bheading{Framework Layer.} 
The Android framework acts as an intermediary layer between applications and the underlying system. It provides APIs that allow apps to interact with hardware sensors, networking, graphics, and user interface components. The framework translates app requests into system calls, enabling communication with binaries, libraries, and the Linux kernel.

\bheading{Application Layer.} 
Apps interact with the Android framework to access system resources like sensors, network, and storage.
The apps include system apps, pre-installed apps, and third-party apps. System apps are essential applications that provide system-level functionalities, while pre-installed apps are the apps provided by the vendors.

\subsection{Android Customization}

Developing an Android-based system involves a multi-stage process that begins with the strategic selection of an Android version as the foundational base image. This critical decision determines the SDK level and technological capabilities available to the system. Following version selection, manufacturers systematically fork the corresponding branch from the Android Open Source Project (AOSP) repositories to establish their development baseline.
The customization phase involves comprehensive system-wide modifications that span both user-space and kernel-space components. Vendors integrate proprietary binaries, services, and core system modifications to achieve their desired functionality. At the kernel level, they implement essential drivers for custom peripherals and make strategic adjustments to other components (e.g., security policies and initialization scripts) to optimize system performance and security.
For VR devices specifically, the extensive array of specialized sensors and peripherals requires substantial kernel-level customization. This includes implementing custom device drivers and developing kernel support infrastructure to enable seamless interaction between userspace APIs and VR-specific hardware components (\eg, eye-tracking sensors).

\begin{table*}[t]
\footnotesize
\centering
\begin{tabular}{rrrrrrr}
\hline
Device & Version & Size & \begin{tabular}[c]{@{}l@{}}Ave. Binary\\ Per Version\end{tabular} & \begin{tabular}[c]{@{}l@{}}Ave. App\\ Per Version\end{tabular} & Date Range & Android version \\ \hline
Quest     & 54  & 100GB & 1956 & 103& June 2019 - December 2023 & Android 7.1.1, 10\\
Quest 2   & 151 & 350GB & 3049 & 117 & September 2020 - December 2024 & Android 10, 12\\
Quest 3   & 37  & 150GB & 3351 & 131& September 2023 - December 2024 & Android 12 \\
Quest Pro & 79  & 170GB & 3290 &127& March 2023 - December 2024 & Android 10, 12\\
Pico Neo 3   & 4   & 14GB  & 4132 &181 & May 2021 - December 2024 & Android 10\\ 
Pico 4    & 4   &  14GB & 4350 & 176 & October 2022 - December 2024 & Android 10\\
\hline
\end{tabular}
\caption{VR Firmware Dataset}
\label{t:dataset}
\vspace{-10pt}
\end{table*}

\subsection{Android Compliance Check}

Android enforces a strict compliance framework to ensure device compatibility and security. Central to this is the Android Compatibility Definition Document (CDD) \cite{android_cdd}, which outlines the technical specifications for {\it Android-compatible} devices. For each release, Google provides a version-specific CDD integrating SDK API documentation and implementation guidelines to ensure consistency. Manufacturers must pass a detailed validation process checking both AOSP compatibility and security to gain certification. The Compatibility Test Suite (CTS)~\cite{cts} is the main assessment tool, testing adherence to CDD standards, ensuring devices meet Android's compatibility and security requirements.

The CDD framework plays a crucial role in balancing innovation with standardization by establishing clear boundaries for vendor customization. While manufacturers retain the flexibility to enhance and differentiate their Android-based products through custom modifications, these adaptations must conform to CDD guidelines to preserve compatibility. 
While Android-based VR devices are not mandated to obtain CDD certification, adherence to these established standards provides a critical framework for evaluating device security posture. For example, the CDD specifications outline essential kernel-level security requirements for specific Android versions, establishing baseline protections that are fundamental to maintaining robust security across VR systems. As there is no such standard for VR devices yet, the requirements from Android CDD serve as valuable reference points for assessing the security maturity of VR implementations.

\subsection{Android Security Features}
Modern Android systems implement a range of security mechanisms across multiple layers of the software stack.
At the kernel level, Android has implemented defenses like kernel page table isolation and KASLR to protect the kernel space. 
Similarly, Android enforces binary hardening techniques~\cite{yut,possemato2021trust} on system binaries such as stack canaries, position-independent executables (PIE), and full RELRO to mitigate various security vulnerabilities.
Moreover, Android relies on SELinux as a mandatory access control framework to confine system processes and limit privilege escalation through fine-grained policies.
These mechanisms form the foundation of Android's defense-in-depth strategy.
Although modern VR devices rely on customized Android versions, they often deviate from the standard Android release process and may include vendor-specific modifications, custom builds, or legacy components.
In this work, we focus on understanding whether these well-established security practices are faithfully translated into VR firmware across device generations or whether gaps exist that could expose VR devices to additional security risks.
We studied existing Android firmware analysis works~\cite{maar2024defects,hernandez2020bigmac,possemato2021trust,gamba2020analysis,hou2022large,hou2023can,sutter2023firmwaredroid,elsabagh2020firmscope,elgharabawy2022sausage,el2021dissecting} and identified critical security features in Android, which motivates our study.  \section{Overview}
\label{sec:overview}

In this section, we first present the research questions guiding our study. We then describe the VR firmware dataset we curated. Finally, we detail the our analysis pipeline used to process and extract relevant information from the firmware. 

\subsection{Research Goals}

As VR platforms continue to grow in complexity and adoption, the firmware powering these devices becomes a critical component in VR systems' security and privacy guarantees.
\rev{Unlike traditional mobile platforms, VR firmware vulnerabilities can compromise both privacy and users’ physical safety. Misplaced Guardian boundaries, camera spoofing, or corrupted sensor data can cause disorientation and injury. Moreover, VR devices process continuous sensor data (e.g., eye/face/body trackers), making kernel-level misconfigurations and weak SEPolicy rules particularly dangerous.}
However, the security of VR firmware and the implications of its design decisions over time remain largely unexplored.
To systematically analyze its security posture, we focus on the following research questions to understand the security of different layers of the firmware stack (\autoref{fig:structure}):

\begin{packeditemize}
    \item \textbf{RQ1:} What kernel-level security features are missing or inconsistently deployed in VR firmware?
    \item \textbf{RQ2:} Are the binaries in VR firmware properly hardened with appropriate defenses?
    \item \textbf{RQ3:} What privileges and security controls are applied to pre-installed applications in VR firmware?
    \item \textbf{RQ4:} How are SELinux policies enforced across firmware versions and device generations?
\end{packeditemize}

We perform a longitudinal analysis across firmware updates and device generations to identify regressions or improvements, and better understand how vendors evolve their security measures for VR devices.

\subsection{VR Firmware Dataset}
Our study focuses on standalone VR headsets running Android-based firmware.
\rev{We compiled a list of major standalone VR vendors and excluded those without publicly available firmware (e.g., HTC-Vive and Varjo).}
To conduct our longitudinal analysis, we curated a comprehensive dataset of firmware images from two leading VR device vendors: Meta and Pico.
\rev{Meta and Pico were selected because their firmware is available, and the two vendors represent over 82\% of the VR headset market~\cite{market-share},
making them highly representative of the consumer VR landscape.}
Specifically, we target the following standalone VR devices: Meta Quest 1, Quest 2, Quest 3/3S, Quest Pro, Pico Neo 3, and Pico 4.

To collect firmware images for these devices, we searched various online sources, including forums, developer communities and archival websites\cite{htcforum,meta_forums,unreal_engine_forums,xda_forums, pico_os,internetarchive2025picoos,cocainetrade2025}.
For Meta Quest devices, we identified a repository~\cite{cocainetrade2025} that hosts a comprehensive archive of firmware updates across all Quest product lines.
For Pico devices, we collected firmware directly from official update channels~\cite{pico_os} and manually retrieved older firmware versions using the Wayback Machine~\cite{internetarchive2025picoos}, which allowed us to access historical download links that were no longer publicly listed.
We  excluded devices that have been available only for a limited period. For instance, Meta Quest 3S was released in October 2024 and the firmwares are only available for a two month duration (Oct - Dec 2024) at the time of our data collection.

Our final dataset includes over $300$ distinct firmware versions across the six VR devices.
The firmware images span a date range from March 2019 to December 2024, and total more than 700GB in size.
\autoref{t:dataset}
summarizes the number of versions per device, their average binary and app counts, and the underlying Android base versions.
We observed a clear progression in the Android base versions across different VR device generations.
The original Quest device initially used Android 7.1.1 as its foundation before transitioning to Android 10 in subsequent updates.
Quest 2 and Quest Pro devices were developed on Android 10 and later received major version updates to Android 12.
The Quest 3 was developed on Android 12, representing the most up-to-date Android adoption in Quest devices. 
In contrast, Pico has maintained Android 10 as the base operating system across their device lineup, demonstrating a more conservative approach to Android version adoption.
This longitudinal dataset provides a strong foundation for analyzing security practices across both vendors and multiple generations of Android-based VR devices.

\begin{figure}[t]
    \centering
    \includegraphics[width=.99\linewidth]{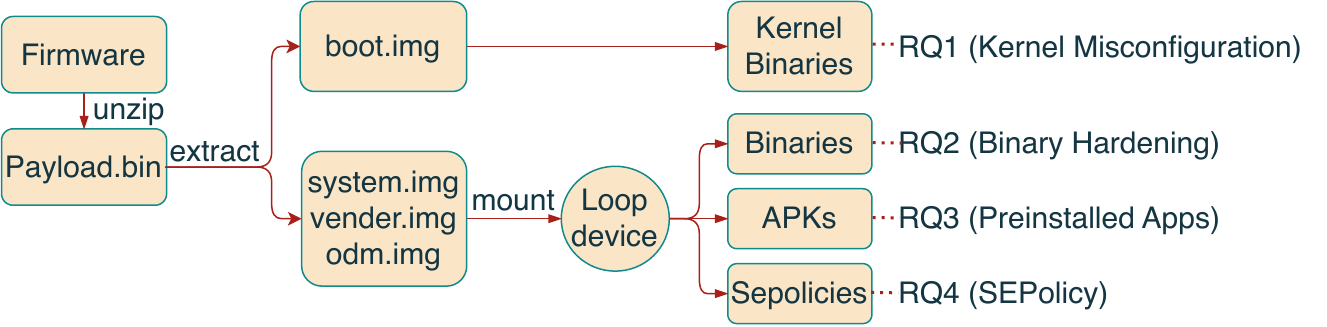}
\caption{High-level Approach}
    \label{fig:workflow}
\end{figure}

\begin{table*}[ht]
\centering
\footnotesize
\begin{tabular}{lll}
\toprule
\textbf{Attack Vector} & \textbf{Mitigation} & \textbf{Building Configuration} \\ \midrule
Stack Overflow & Stack Protector & ** \cc{CONFIG\_HAVE\_STACKPROTECTOR} or \cc{CONFIG\_STACKPROTECTOR\{\_STRONG\}}\\
Control Flow Hijacking & KASLR & * \cc{CONFIG\_RANDOMIZE\_BASE} \\
Heap Corruption & Freelist Randomization & \cc{CONFIG\_SLAB\_FREELIST\_RANDOM} \\
Information Leakage & USERCOPY & ** \cc{CONFIG\_HARDENED\_USERCOPY} \\
Buffer Overflow & Fortify Source & \cc{CONFIG\{\_ARCH\_HAS\}\_FORTIFY\_SOURCE} \\
Code Injection & Non-executable Memory & ** \cc{CONFIG\{\_ARCH\_HAS\}\_STRICT\_KERNEL\_RWX} or \cc{CONFIG\_DEBUG\_RODATA} \\ 
Privilege Escalation & Restrict Userspace Memory Access
& ** \cc{CONFIG\_CPU\_SW\_DOMAIN\_PAN} or \cc{CONFIG\_ARM64\_SW\_TTBR0\_PAN} \\
Meltdown & Kernel Page Table Isolation & ** \cc{CONFIG\_UNMAP\_KERNEL\_AT\_EL0} \\
Spectre & Branch History Overwrite & \cc{CONFIG\_MITIGATE\_SPECTRE\_BRANCH\_HISTORY} \\
Code reuse attacks & Enable CFI in kernel & * \cc{CONFIG\_CFI\_CLANG} {\bf and} \cc{CONFIG\_SHADOW\_CALL\_STACK} \\
Data Leaks & Prevent use of uninitialized local variables & $^+$ \cc{CONFIG\_INIT\_STACK\_ALL\{\_ZERO\}} \\
Data Leaks from Heap & Enable Heap Initialization & $^+$ \cc{CONFIG\_INIT\_ON\_ALLOC\_DEFAULT\_ON} \\
Use After Free Vulnerability & Sanitization & \cc{CONFIG\_DEBUG\_LIST} \\
Speculative Execution & BPF JIT enabled & \cc{CONFIG\_BPF\_JIT\_ALWAYS\_ON} \\
Free list exploits & Harden Slab allocator & \cc{CONFIG\_SLAB\_FREELIST\_HARDENED} \\
Stack corruption & Virtually mapped kernel pages with guard pages & \cc{CONFIG\_VMAP\_STACK} \\
Type confusion & Hardware enforced instructions & \cc{CONFIG\_ARM64\_UAO} \\
\bottomrule
\end{tabular}\caption{Attack vectors and mitigations in the kernel covered in our study; X\{Y\} means that either X or XY can be used for mitigation. ** means MUST; * means STRONGLY RECOMMENDED in Android 10; + means STRONGLY RECOMMENDED added in Android 12. For the configuration flags without any marks, they are the ones SUGGESTED by~\cite{maar2024defects}.}
\label{t:attack_vectors}
\vspace{-12pt}
\end{table*}

\subsection{Firmware Image Analysis}

\autoref{fig:workflow} illustrates our high-level workflow used to extract, process, and analyze artifacts from each VR firmware image.
Our goal is to retrieve kernel binaries, user-space executables, pre-installed applications, and SELinux policies from each image to support the security analyses defined by our research questions.

We analyze each firmware image via a three-phase approach.
First, we decompress the firmware packages.
VR firmware images are typically distributed as \texttt{.zip} archives.
We use standard tools such as \cc{unzip}  to extract the contained payloads (i.e., \cc{payload.bin} and Brotli compressed files). 
Next, we extract the relevant system partition images.
For Meta firmware, we use a payload extraction tool~\cite{payload-dump} to extract key image files, including \cc{boot.img}, \cc{system.img}, \cc{vendor.img}, and \cc{odm.img}.
For Pico firmware, we use \cc{brotli}~\cite{Brotli} to retrive the decompressed Brotli (\cc{.br}) files which contain these partition images.
Lastly, we mount these images to extract the relevant artifacts for our analysis using standard Linux loop device and extraction tools (e.g., \cc{boot.img} editor~\cite{boot-extractor}).
Specifically, we extract the kernel binaries from \cc{boot.img} for RQ1 and extract native binaries, pre-installed applications, and SELinux policies from \cc{system.img}, \cc{vendor.img}, and \cc{odm.img} one by one for answering RQ2, RQ3 and RQ4.
To identify the Android version, we parse the \cc{build.prop} configuration file located in the mounted \cc{system} partition.
For Meta devices, we use the \cc{ro.build.id} field, while for Pico devices we rely on \cc{ro.system.build.version.release} and \cc{ro.system.build.version.sdk}.
\autoref{t:system_paths} (in Appendix) summarizes the key paths and properties we use to extract security-relevant data from each firmware image.
This pipeline enables scalable, automated analysis 
across hundreds of firmware versions and serves as the foundation for the longitudinal firmware analysis presented in the remainder of the paper.

\section{RQ1: Kernel Misconfiguration}
\label{s:rq1}

\subsection{Motivation}
The Android Compatibility Definition Document (CDD) \cite{android_cdd} establishes critical security requirements for kernel compilation, mandating specific security features to mitigate potential vulnerabilities. Non-compliance with these requirements can lead to severe security implications, as demonstrated by CVE-2018-9568 \cite{CVE-9568}, which enabled privilege escalation attacks on Android systems through kernel misconfiguration. Given that VR devices utilize customized Android kernels, they inherit similar vulnerability patterns.
As one concrete example, CVE-2018-9568 was successfully exploited on Oculus Quest devices~\cite{Github-9568} for privilege escalation, illustrating how security weaknesses can propagate across the Android ecosystem to VR platforms.

In response to such vulnerabilities, the Android Open Source Project (AOSP) has implemented robust kernel-level security mitigations \cite{Google-blog}, such as Kernel Address Space Layout Randomization (KASLR) and Privileged Access Never (PAN). 
However, prior research works~\cite{zhang2021investigation} have shown that nearly half of the CVEs are patched on OEM devices roughly 200 days or more after the initial patch is publicly committed in the upstream.
This significant patch deployment latency, combined with VR devices' shared kernel architecture with Android, underscores the critical importance of systematic kernel security misconfiguration analysis in VR platforms. Such analysis is essential to ensure VR devices maintain parity with Android's evolving security standards.

\subsection{Methodology}
For identifying the mitigations present in the kernel, we need to extract the kernel configuration file (\cc{.config})  from the kernel binary. 
To do so, we used the \textit{extract-ikconfig} script \cite{extract-ikconfig}. 
Unlike existing studies that face difficulties in extracting kernel configurations~\cite{yut,possemato2021trust},
we were able to extract all the \cc{.config} files from the kernel binaries because all VR firmware were compiled with the \cc{CONFIG\_IKCONFIG} flag~\cite{ikconfig_lkddb};
This config flag is a kernel configuration option in the Linux kernel that allows the kernel to include its own configuration file as part of the compiled kernel image, and it has been required since Android 8.0. 
As most of the VR firmware in our dataset uses an Android version higher than 8.0 and the ones that have used Android 7.1.1 (from Quest) also have this flag enabled, we were able to extract kernel configurations successfully from all the kernels we collected. 

To examine kernel misconfigurations, we first manually reviewed the Android CDD and extracted the relevant kernel-level security mitigations. The CDD provides a set of mandatory kernel mitigations, which focus on addressing only a subset of security issues. However, these categories do not fully encompass the range of potential vulnerabilities that could compromise system security; therefore, we have examined prior works~\cite{maar2024defects} and the additional mitigations that increase the security posture of the kernel. These mitigations and associated flags are present in~\autoref{t:attack_vectors}. 
In total, we check $17$ mitigations.
The mitigations can be classified into three types - MUST (from CDD), STRONGLY RECOMMENDED (from CDD), and SUGGESTED (from \cite{maar2024defects}). 
Devices are expected to implement those which are MUST, and STRONGLY RECOMMENDED at least for a better security posture.
Then, we have developed a script by taking the requirements as reference to analyze the kernel configuration (the extracted \cc{.config}  file) from the VR kernel binaries.
We have examined all mandatory and suggested kernel-level mitigations  to measure their adoption rates among VR devices. \looseness=-1

\begin{figure}[t]
    \centering
    \includegraphics[width=0.9\columnwidth]{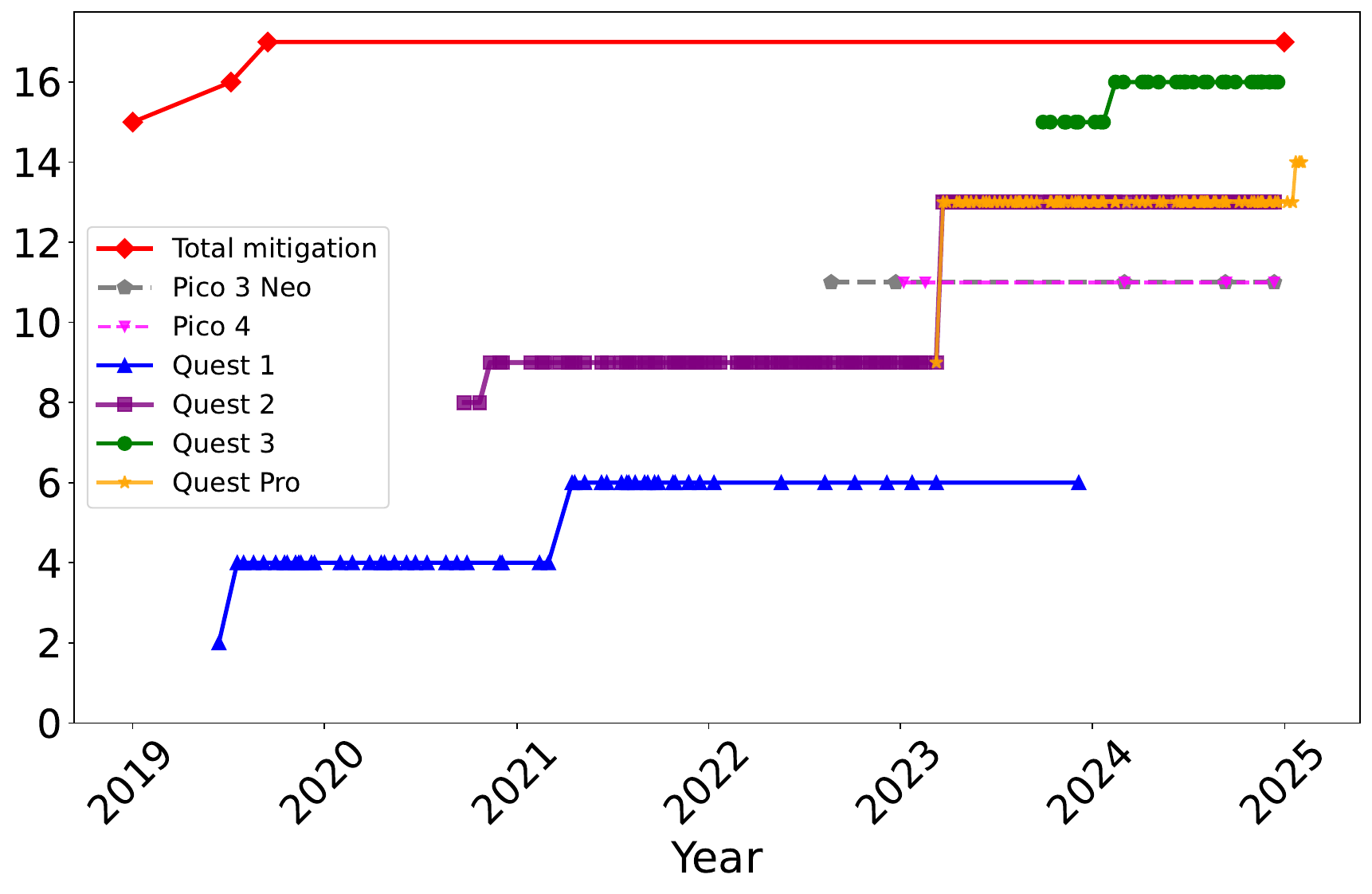}
    \vspace{-12pt}
    \caption{Kernel mitigations adopted by the VR devices; for Pico devices we only have four versions of each.}
    \label{fig:RQ1}
\end{figure}

\subsection{Longitudinal analysis}

In the VR firmware dataset we investigated, 
we observed that none of the firmwares adopted all the available mitigations. 
The mitigations adopted by different devices are shown in~\autoref{fig:RQ1}.
Quest 1 began with only two kernel-level mitigations in mid-2019 and then was increased to six by early 2021, but it received no further enhancements despite additional mitigations becoming available. 
Quest 2 was launched in late 2020 with eight mitigations and reached nine within a few months, showing a similarly stagnant update pattern afterward. 
From the graph, we observe that the Quest Pro adopts the same kernel-level mitigations as Quest 2, and the updates were released on similar days. 
In contrast, newer models like Quest 3 adopted a significantly larger subset of available mitigations. 
The jumps in adopting kernel-level configurations in Quest, Quest 2 and Quest Pro are observed because of updates of Android versions, indicating that vendors would only include kernel-level mitigations during major Android version updates.
In the case of the Pico 4 and Pico 3 Neo, we observed that the kernel version remains unchanged, along with the adoption of mitigations and there is no positive trend in adopting newer mitigations across the 4 versions we investigated. 

Overall we find that Quest 3 has good security posture compared to other devices.
On the positive side, more mandatory mitigations were adopted by Quest devices over time, even though they were not included in the initial versions. 
Although vendors may not always upgrade to the latest kernel versions, they add mitigations longitudinally through the device life cycle to enhance security over time. 
This approach, while improving security to some extent, does not fully mitigate the risks associated with outdated kernel versions.
We also studied  the missing kernel mitigations that were not adopted in the latest Quest and Pico devices in our dataset, shown in~\autoref{t:missing-kernel-configs}.
We observe that both Quest and Pico devices miss many mitigations. 

\bheading{Kernel versions.}
One observation from our analysis is that VR devices tend to use older LTS (Long-Term Support) kernel versions than the latest LTS versions, the slower pace of Linux kernel adoption could explain the absence of certain kernel-level mitigations. 
The kernel version associated each device when they were released, along with the latest LTS version available, are shown in~\autoref{t:missing-kernel-lts}.
We can see that the Quest, released in May 2019, adopted Linux kernel version 4.4, originally released in 2016, even though it could have adopted version 4.19, released in October 2018. 
Similarly, the Quest 2, launched in October 2020, continued
to use Linux kernel version 4.19, despite the availability of version 5.4, released in 2019. 
Quest Pro was released in 2022, but it still adopted kernel version 4.19. 
The Quest 3, released in 2023, adopted Linux kernel version 5.10, even though version 6.1 had been available since 2022. 
All Pico headsets use the outdated Linux kernel version 4.19. Even the Pico 4, released in 2024, continued to rely on Linux 4.19, highlighting Pico’s significant delay in adopting newer kernel versions. 
One possible reason vendors continue to use these older kernels is their focus on stability, as introducing newer kernels may cause issues that take time to resolve and require significant engineering efforts. 
However, this approach creates security gaps, as these devices are not adopting newer kernel-level mitigations.

\begin{table}[]
\footnotesize
\begin{tabular}{rrrrrr}
\toprule
Release date & Device    & Kernel Version  & Latest LTS  & Android Version\\ \midrule
May 2019     & Quest     & 4.4.x    & 4.19.x     &  7.1.1 \\
October 2020 & Quest 2   & 4.19.x   & 5.4.x      &  10\\
October 2022 & Quest Pro & 4.19.x   & 5.15.x     &  10\\
October 2023 & Quest 3   & 5.10.x   & 6.1.x      &  12 \\ May 2021     & Pico 3 Neo & 4.19.x  & 5.10.x     &  10 \\ 
October 2022 & Pico 4    & 4.19.x   & 6.6.x      &  10 \\
\bottomrule
\end{tabular}
\caption{Kernel versions used at the release time of each device and the latest LTS version available at that time.}
\label{t:missing-kernel-lts}
\vspace{-10pt}
\end{table}

\begin{table*}[t]
\centering
\footnotesize
\begin{tabular}{rrl}
\toprule
\textbf{Device}    & \textbf{Kernel Version}  & \textbf{Missing Kernel Mitigations} \\ 
\midrule
Quest              & 4.4.x    & \makecell[l]{*\cc{CONFIG\_CFI\_CLANG, CONFIG\_SHADOW\_CALL\_STACK, CONFIG\_SLAB\_FREELIST\_RANDOM,}\\
                                \cc{CONFIG\_FORTIFY\_SOURCE, CONFIG\_MITIGATE\_SPECTRE\_BRANCH\_HISTORY, CONFIG\_DEBUG\_LIST,}\\
                                \cc{CONFIG\_BPF\_JIT\_ALWAYS\_ON, CONFIG\_SLAB\_FREELIST\_HARDENED, CONFIG\_VMAP\_STACK, CONFIG\_ARM64\_UAO}} \\\midrule
Quest 2 / Quest Pro & 4.19.x   & **\cc{CONFIG\_UNMAP\_KERNEL\_AT\_EL0,} $^{+}$ \cc{CONFIG\_INIT\_STACK\_ALL\_ZERO,} $^{+}$\cc{CONFIG\_INIT\_ON\_ALLOC\_DEFAULT\_ON,} *\cc{CONFIG\_CFI\_CLANG} \\\midrule
Quest 3            & 5.10.x   & $^{+}$\cc{CONFIG\_INIT\_ON\_ALLOC\_DEFAULT\_ON} \\\midrule
Pico 3 / Pico 4    & 4.19.x   & \makecell[l]{**\cc{CONFIG\_ARM64\_SW\_TTBR0\_PAN,} **\cc{CONFIG\_UNMAP\_KERNEL\_AT\_EL0, CONFIG\_MITIGATE\_SPECTRE\_BRANCH\_HISTORY,}\\
                                *\cc{CONFIG\_CFI\_CLANG, CONFIG\_DEBUG\_LIST, CONFIG\_BPF\_JIT\_ALWAYS\_ON}} \\
\bottomrule
\end{tabular}
\caption{Missing kernel mitigations in latest firmware versions.
}
\label{t:missing-kernel-configs}
\vspace{-10pt}
\end{table*}

\begin{mdframed}[
  roundcorner=5pt,   linewidth=0.8pt,   linecolor=black,   backgroundcolor=gray!10,
  innertopmargin=4pt,innerbottommargin=4pt,
  innerleftmargin=4pt,
  innerrightmargin=4pt
]
\textbf{Summary.}  
VR devices demonstrate issues related to kernel misconfiguration. Meta Quest devices are striving to enhance the implementation of mitigations at the kernel level, while, in contrast, Pico devices show no evident longitudinal change in the adoption of kernel-level mitigations. Moreover, these devices are not proactively adopting the most recent stable Linux versions available. This delayed adoption exacerbates security vulnerabilities, as older kernels are deficient in incorporating newer mitigations vital for countering emerging threats.
\end{mdframed}

\section{RQ2: Binary Hardening}
\label{s:rq2}

\subsection{Motivation}
Binary hardening~\cite{wikipediaHardeningcomputing} is a security technique that focuses on analyzing and modifying binary executable to safeguard against common exploits. 
This approach often involves non-deterministically altering control flow and instruction addresses, making it difficult for attackers to reuse program code and execute successful exploits. 
Binary compliance refers to whether a binary follows the hardening techniques prescribed by Android to prevent security issues. 
Lack of binary compliance can introduce vulnerabilities in the VR systems. This issue is demonstrated by incidents like the Stagefright vulnerability (CVE-2015-3864) \cite{CVE-2015-3864}  that exploited the multimedia framework binary that allows remote attackers to execute arbitrary code. \looseness=-1

Although Android devices have increasingly adopted advanced binary hardening techniques~\cite{possemato2021trust}, it remains unclear to what extent VR platforms, such as Meta Quest and Pico, implement these measures. 
In this section, our objective is to systematically investigate the adoption of binary hardening techniques on VR devices to identify potential inconsistencies and gaps. 
Inconsistent implementation may expose systems to vulnerabilities that might otherwise be avoided. For instance, Meta's application of Relocation Read-Only (RELRO) in Quest 2~\cite{quest-blog} demonstrates how vital effective binary hardening is in ensuring the security of VR ecosystems.

\begin{figure*}
  \centering
  \begin{minipage}[t]{0.32\textwidth}
    \centering
    \includegraphics[width=\linewidth]{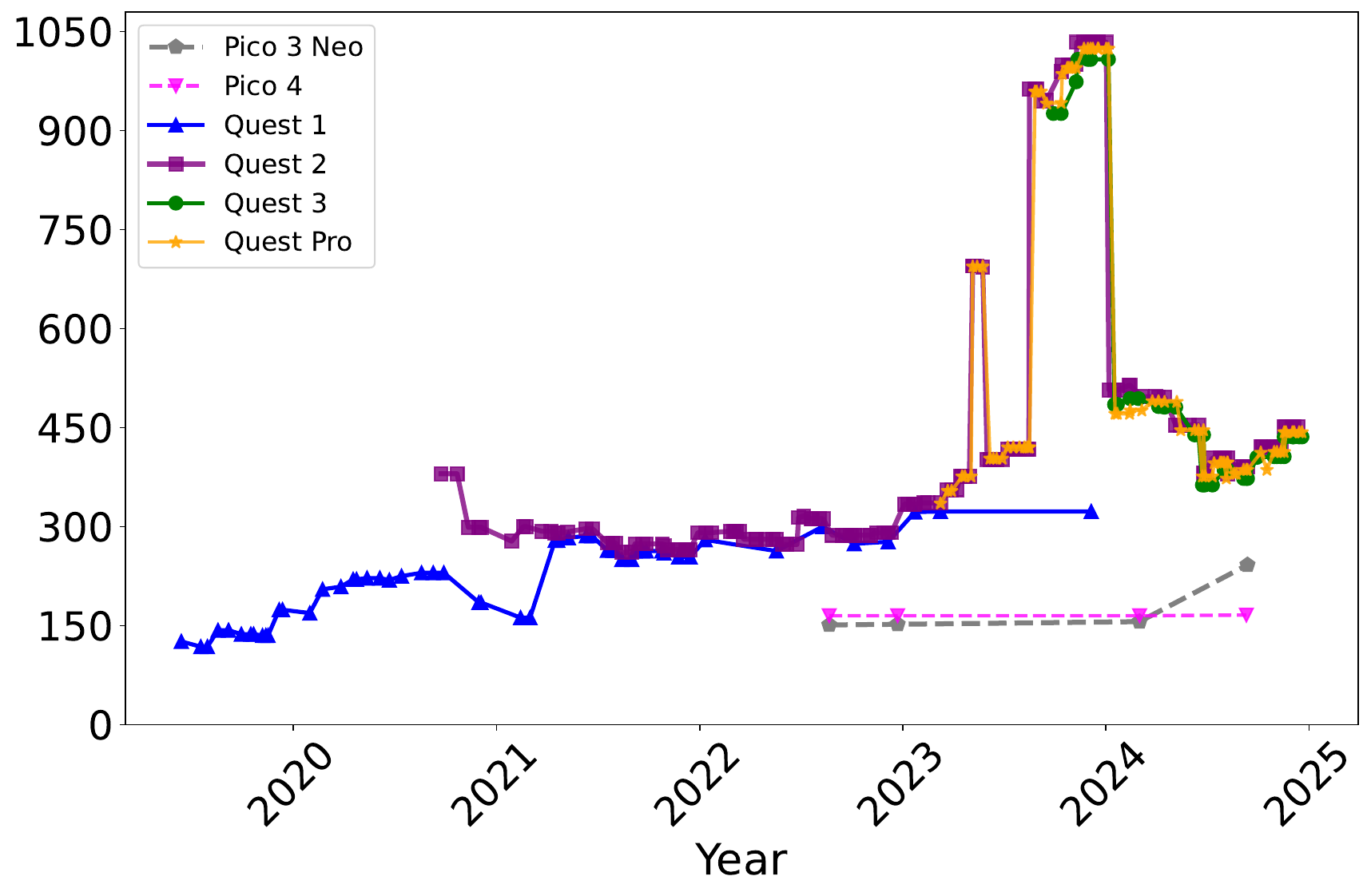}
    \vspace{-20pt}
    \caption{Binaries without Canaries}
    \label{fig:canary}
\end{minipage}\hfill
  \begin{minipage}[t]{0.32\textwidth}
    \centering
    \includegraphics[width=\linewidth]{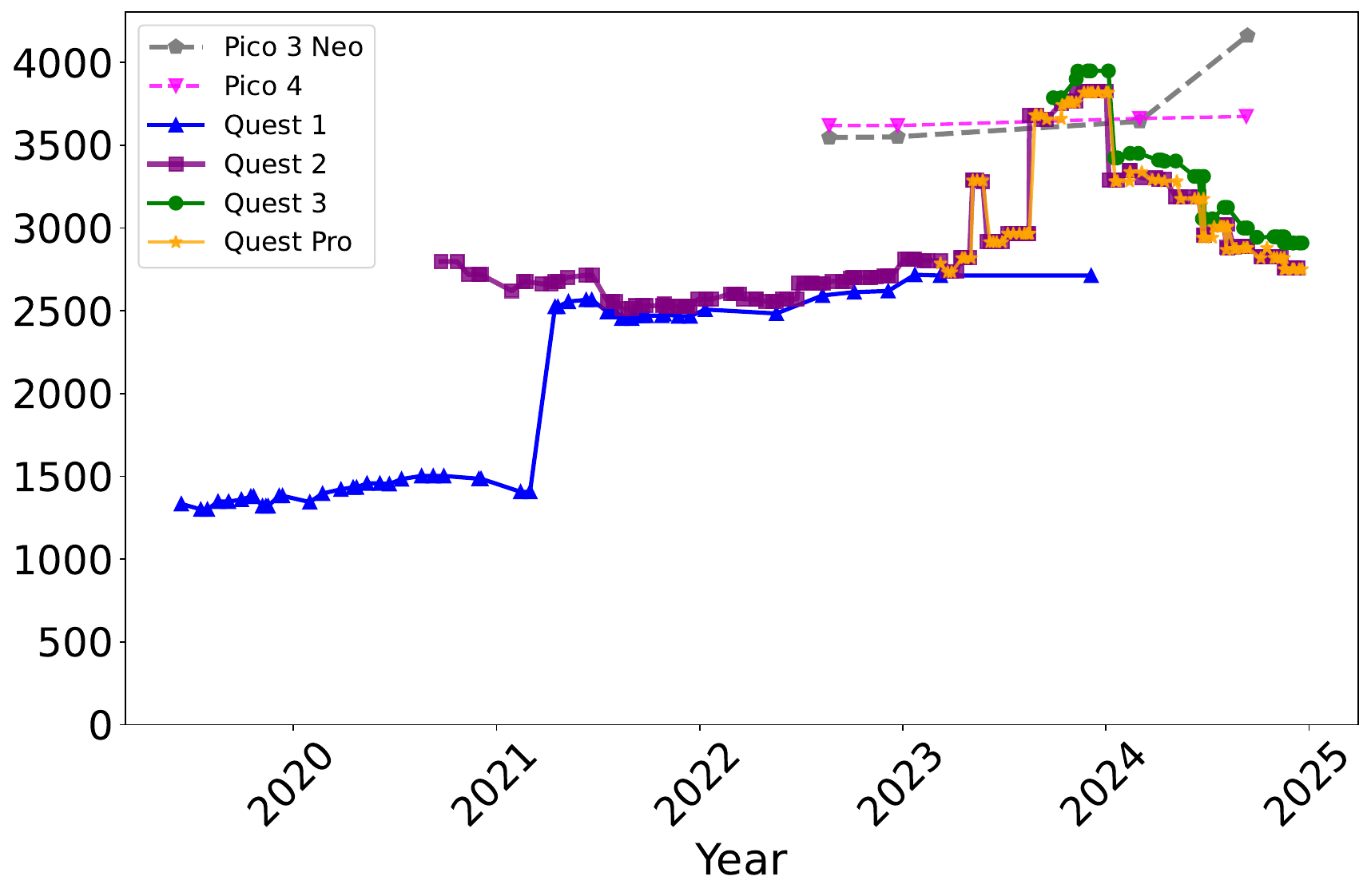}
    \vspace{-20pt}
        \caption{Binaries without CFI}
    \label{fig:CFI}
  \end{minipage}\hfill
  \begin{minipage}[t]{0.32\textwidth}
    \centering
    \includegraphics[width=\linewidth]{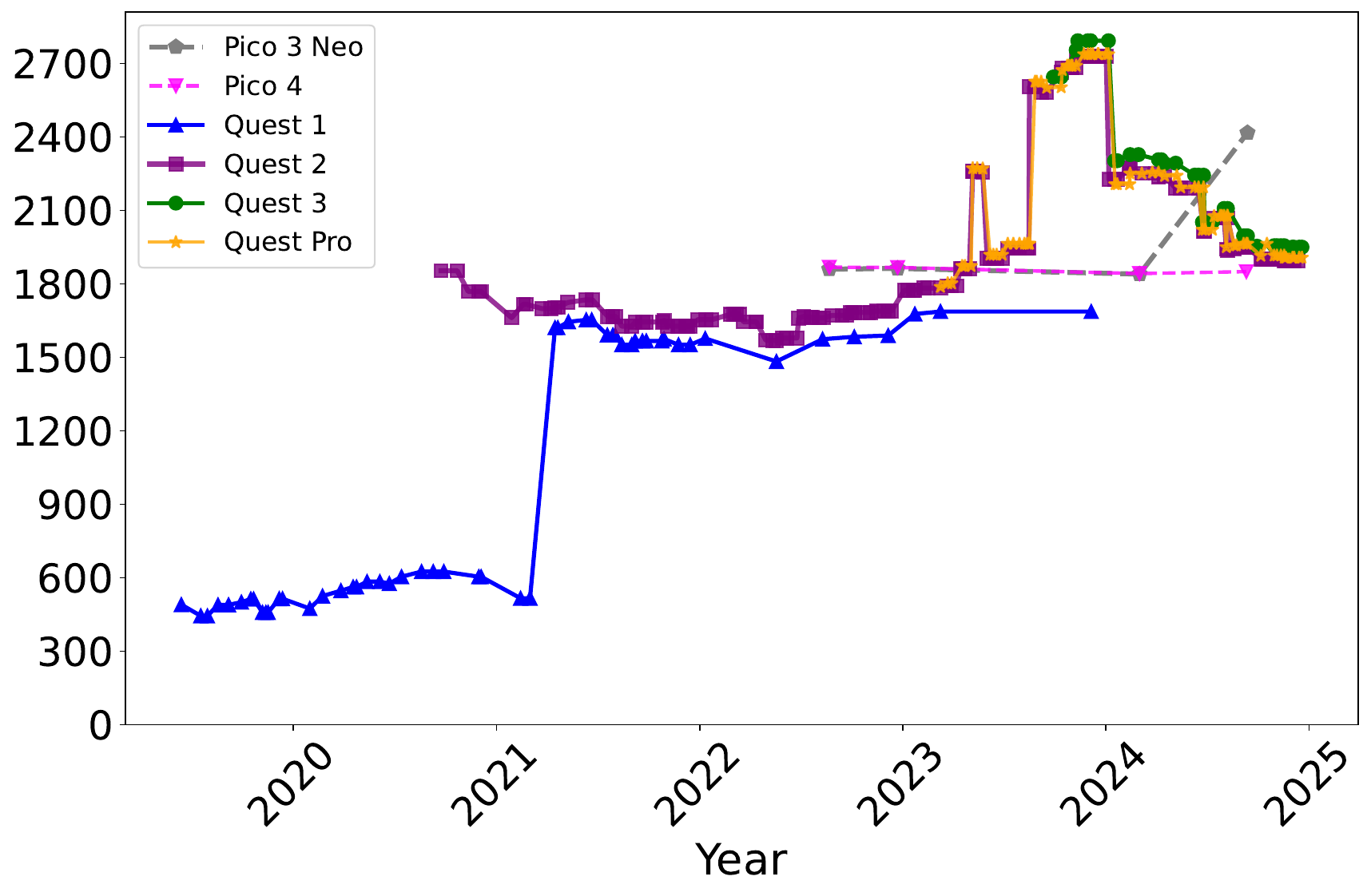}
    \vspace{-20pt}
    \caption{Binaries without Fortify Source}
    \label{fig:FORTIFY}
  \end{minipage}
\end{figure*}

\begin{figure}[t]
    \centering
    \includegraphics[width=0.8\columnwidth]{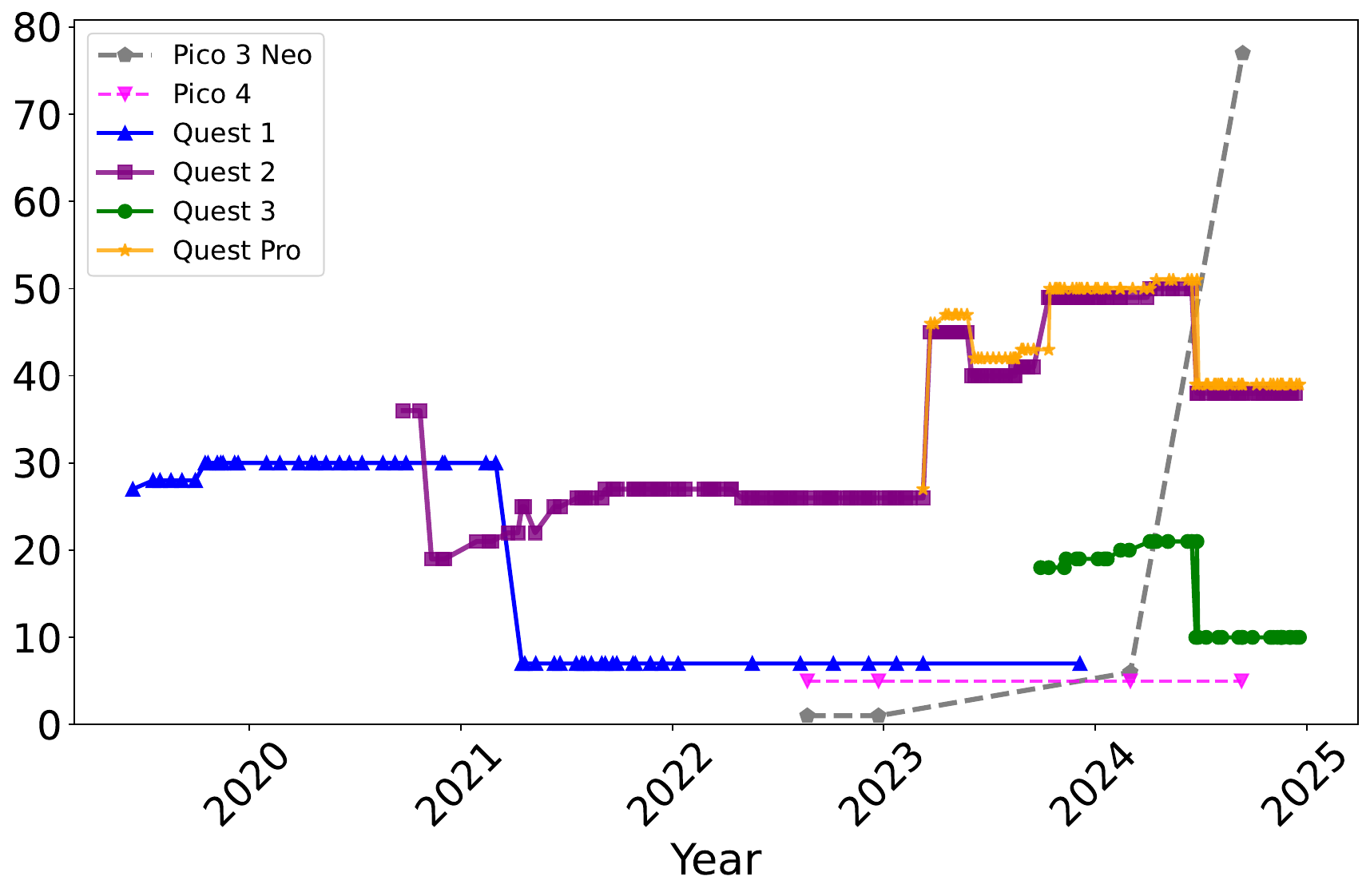}
    \vspace{-12pt}
    \caption{Binaries without NX}
    \label{fig:NX}
    \vspace{-10pt}
\end{figure}

\begin{figure}[t]
    \centering
    \includegraphics[width=0.8\columnwidth]{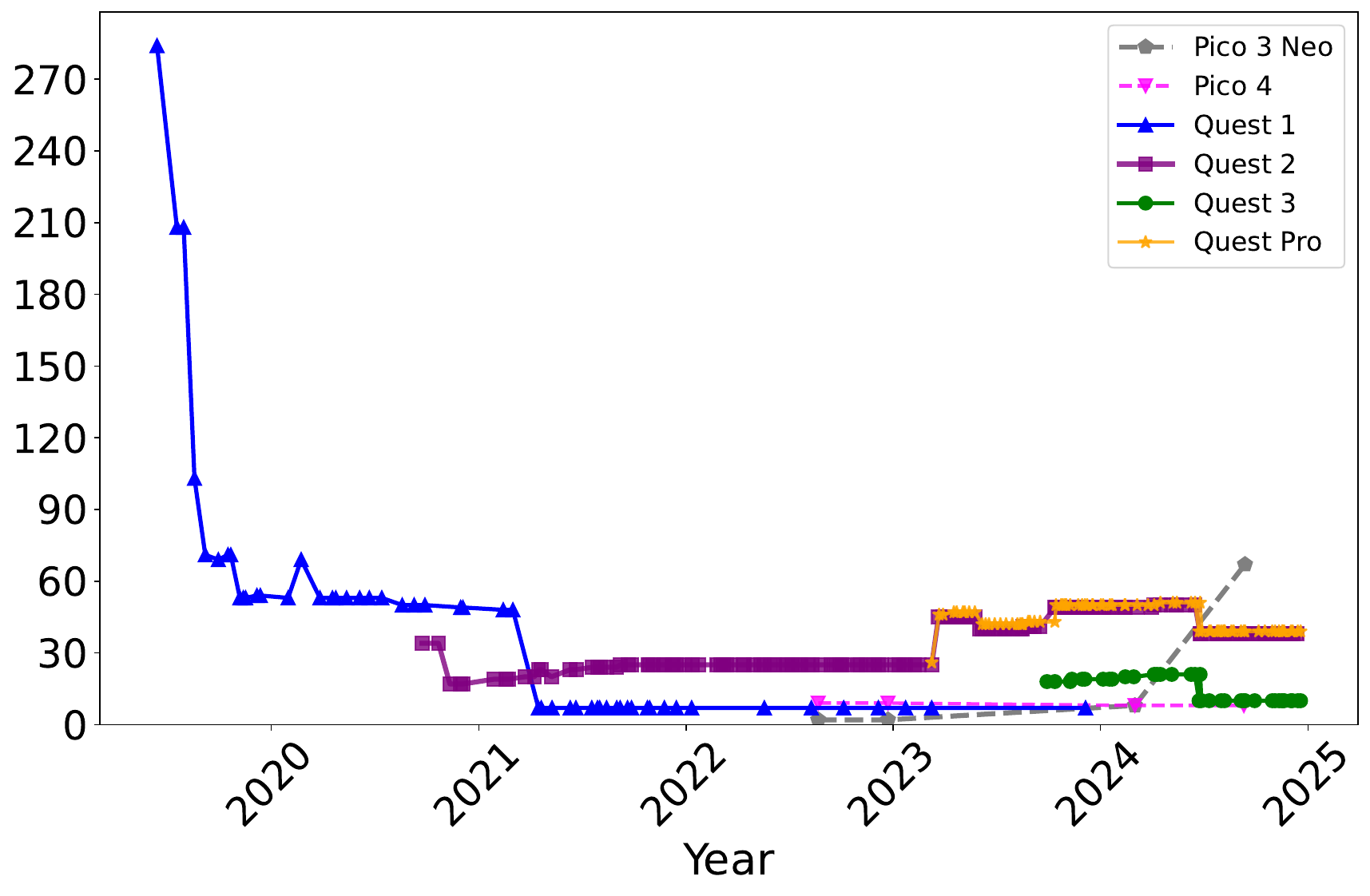}
    \vspace{-12pt}
    \caption{Binaries without RELRO}
    \label{fig:RELRO}
\end{figure}

\subsection{Methodology}
Our study focused on five specific mitigation techniques that should always be enabled according to CDD: Stack Canaries, Full Relocation Read-Only (Full RELRO), Fortify Source, Control Flow Integrity (CFI), and the NX bit,
These techniques collectively work to defend against prevalent exploits and minimize potential vulnerabilities.
Our approach to identifying these mitigations was adapted from methodologies outlined in past research~\cite{yut,possemato2021trust}, which primarily explored these mitigations within embedded systems and Android environments, which we detail below.

\begin{packeditemize}
  \item \textbf{Stack Canaries} are a defense mechanism against stack-based buffer overflow vulnerabilities by detecting any stack corruption before the function returns. To verify the implementation of Stack Canaries, we check \cc{\_\_stack\_chk\_fail} within symbols or the Global Offset Table (GOT) table.
 
    \item \textbf{Control Flow Integrity (CFI)} protects against attacks such as Return-Oriented Programming (ROP) and Jump-Oriented Programming (JOP) by maintaining the correct control flow. We verify the binary's symbol table for the existence of \cc{.cfi} symbols. \looseness=-1
    
    \item \textbf{Fortify Source} aims to improve memory safety by identifying potentially unsafe uses of standard library functions during the compilation process. Detection involves searching for \cc{*\_chk} functions within the binary; additional indicators include strings like \cc{*buffer} \cc{overflow} \cc{detected*}, which signify binary fortification. \looseness=-1

    \item \textbf{NX bit} is a defense against buffer overflow attacks by designating certain memory regions as non-executable, thereby preventing code execution in those areas. We inspected the binary for the \cc{PT\_GNU\_STACK} segment. If found and marked non-executable, it signifies that the binary seeks a Non-executable Stack.
    
    \item \textbf{Full RELRO} protects critical relocation sections of a binary by rendering them read-only, thereby preventing GOT overwrite attacks. The absence of a \cc{GNU\_RELRO} segment indicates that the binary lacks RELRO protection.
    
\end{packeditemize}

\subsection{Longitudinal analysis}

\bheading{Binaries without Canaries, CFI, and Fortify Source.}
The trend of binaries without canaries is shown in~\autoref{fig:canary}.
Quest 1 had around 150 system binaries without canaries in 2019, gradually increases to about 300 by 2023, and then remains flat.
Quest 2 enters the market in late 2020 with around 350 binaries without canaries, stays relatively stable at about 300 through 2022-2023, then increases rapidly in 2023-2024.
Quest 3 and Quest Pro  both show similar patterns (overlapping).
All headsets except Quest 1 show a dramatic spike reaching peaks of around 1000 binaries without canaries in early 2024.
After this spike, there's a significant drop, followed by a period of stabilization through 2025.
The Quest 1 appears to be the only Quest headset that did not experience the dramatic spike in late 2023.
Pico 4 did not experience any noticeable changes,
while Pico 3 Neo had an increase in mid 2024.
The  CFI (\autoref{fig:CFI}) and Fortify Source (\autoref{fig:FORTIFY}) show a similar trend, albeit the detailed numbers are different.

\bheading{Binaries without NX.}
As observed in \autoref{fig:NX}, Quest 1 maintains a consistent level of binaries without NX until the start of 2021, after which there is a notable drop, then it fluctuates through 2024. This suggests a significant enhancement for enforcing NX in 2021, coinciding with an Android version update. 
Since 2021, Quest 2's count of binaries without NX remained relatively stable but saw a spike in 2023 due to an Android upgrade, 
followed by fluctuations into 2024. 
Quest Pro exhibits a trend similar to Quest 2, though with slightly elevated values at some points in 2023. Quest 3 begins with strong NX protection (indicated by a low number of binaries without NX), and by 2024, this count is nearly halved.
Pico 4 does not show any change, while Pico 3 Neo has a significant jump of binaries without NX in mid 2024.

\bheading{Binaries without RELRO.}
As observed in \autoref{fig:RELRO}, Quest 1  starts extremely high at around 280 binaries with no RELRO protection in early 2019, then shows a dramatic decline over 2019, stabilizing around 50-60 through 2020. In early 2021, it drops sharply again to near zero and remains at that minimal level through 2023.  
Quest 2 starts in late 2020 with a count much lower than Quest 1's initial values.
It maintains this level through 2022-2023,
then fluctuates but generally maintaining this range through 2024.
Quest Pro shows the same trend as Quest 2.
Quest 3 starts in late 2023 with the lowest initial values of any headset (around 20), suggesting it launched with better RELRO protection from the start, and then drops to even lower levels in 2024.
This shows Meta's progressive approach in enhancing the RELRO protection in Quest devices.
Similar to other results of Pico devices, Pico 4 does not show any change, while Pico 3 Neo has a significant jump in mid 2024.

\bheading{Spikes.}
We observed noticeable spikes in the number of not hardened binaries for Quest 2 and Quest Pro after 2023 (\autoref{fig:canary},~\autoref{fig:CFI},~\autoref{fig:FORTIFY}).
The increase was due to the addition of binaries in preinstalled apps, most of which lacked hardening. All the added binaries had names starting with ``libxplat''. In subsequent updates, many of these binaries were removed, specifically the same ones that had been previously added. Upon investigation, we found that these binaries were intended to provide cross-platform functionality. 
We conjecture that vendors might have opted out of these hardening techniques due to performance reasons~\cite{dang2015performance}, as using these hardening techniques may penalize performance. 

\bheading{Latest version.}
We analyze the latest versions in~\autoref{sec:app:rq2}.

\begin{mdframed}[
  roundcorner=5pt,   linewidth=0.8pt,   linecolor=black,   backgroundcolor=gray!10,
  innertopmargin=4pt,innerbottommargin=4pt,
  innerleftmargin=4pt,
  innerrightmargin=4pt
]
\textbf{Summary.} Within the VR ecosystem, vendors frequently neglect binary hardening techniques, particularly during major Android version upgrades. While longitudinally we see a slow but steady reduction in binaries lacking such security measures, the consistent absence of CFI and Fortify Source across thousands of binaries reveals a significant gap in binary hardening in VR devices.
\end{mdframed}
 
\section{RQ3: Preinstalled Apps}
\label{s:rq3}

\subsection{Motivation}
Preinstalled apps are those that come with the firmware; the user cannot uninstall these apps without root.
Previous studies~\cite{gamba2020analysis,elsabagh2020firmscope,guo2024empirical} have identified the presence of security or privacy-related issues in pre-installed apps, such as privilege escalation, command execution, etc.
For example, CVE-2018-9525~\cite{CVE-2018-9525} demonstrated privilege escalation due to permission bypass;
a broadcast receiver did not enforce \cc{android.permission.MANAGE\_SLICE\_PERMISSIONS},
which caused privilege escalation. 
In this section, we first study the ecosystem of the VR system apps, analyze app manifests, and check how the permission usage and custom permission declaration change across the firmware versions in the preinstalled apps.

\subsection{Methodology}

We study two aspects of preinstalled apps on VR devices: security flags in manifest files, and permissions.

\subsubsection{Security flags in manifest files.} 
The \cc{AndroidManifest.xml} file is like a blueprint for an app, it declares essential information such as the app’s package name, components, required permissions, hardware features, and the minimum Android version it supports. The system uses it to understand how to launch and run the app.
For investigating the manifest of system apps, we use Androguard~\cite{androguard} to decompile the \cc{apk} file and develop a custom script to check the presence of security flags in the manifest of the apk file.
We compile the following list of security flags based on prior work~\cite{guo2024empirical}:
\begin{packeditemize}
    \item \cc{allow\_backup} specifies whether the app's data can be backed up and restored via Android's backup mechanism. Setting the \cc{allow\_backup}  flag to \cc{true} allows user data to be stored in the cloud. However, this may result in the unintended exposure of sensitive information, particularly if the application stores sensitive user or private data.
    \item \cc{debuggable}  indicates whether an application can be debugged. Typically, this flag is set to true in debug builds and false in release builds. If this flag is set to \cc{true} in a release build, attackers can leverage debugging tools to inspect the internal state of the application, examine variable values, and potentially uncover sensitive information or vulnerabilities.
    \item \cc{use\_cleartext\_traffic} indicates whether an application allows plaintext (unencrypted) HTTP traffic. Setting this flag to \cc{true} allows the application to use plain-text HTTP connections. This makes the app susceptible to man-in-the-middle (MITM) attacks, as unencrypted data transmitted over the network can be intercepted and altered.\end{packeditemize}

\subsubsection{Permission analysis.}
We have developed a custom script based on Androguard~\cite{androguard} to extract all the permissions used along with their protection levels from the framework-level apk files (\cc{com.oculus.os.platform-res.apk}, \cc{horizonos.platform-res.apk} and \cc{framework-res.apk}, which define system-wide permissions) as well as preinstalled apps (which define custom permissions).
We further study three aspects of permission issues:
1) permission categorization based on app types;
2) change of permission protection level; and
3) permission inconsistencies.

\bheading{Permission categorization based on app types.}
First, we classify the apps into three categories.
{\bf User-launchable apps} have a user interface that can be launched by users. We parsed the manifest file to check the presence of \cc{android.intent.action.MAIN} and 
\cc{android.intent.category.LAUNCHER}.
{\bf Android system apps} are part of the Android system. We identified these apps by examining package names that follow the format \cc{com.android.*}.
{\bf Vendor specific apps} are the remaining apps, including the vendor-specific ones (e.g., \cc{com.oculus.*}).
Together, these three categories encompass all preinstalled applications.

\bheading{Change of permission protection levels.}
Once the applications and their respective protection levels were identified, we marked the application permissions that displayed varying protection levels across device firmware updates. This suggests that the privileges associated with specific permissions were either enhanced or reduced for certain reasons.
For instance, if a permission initially set to \cc{normal} is later elevated to \cc{system}, this suggests that there may have been an event necessitating tighter control.

\bheading{Permission inconsistencies.}
We have further analyzed two kinds of inconsistent permission scenarios:
{\it Residual Permission:} permissions that are declared in the system but not used by any preinstalled apps; and
{\it Phantom Permission:} permissions that are used by preinstalled apps but not declared in the system. 
In both instances, the permission settings are improperly configured.

\subsection{Longitudinal Analysis}

\begin{table}[t]
\centering
\footnotesize
\setlength{\tabcolsep}{3pt}
\begin{tabular}{rccc}
\hline
\textbf{Device} & \textbf{use\_cleartext\_traffic} & \textbf{allow\_backup} & \textbf{debuggable} \\
\hline
Quest     & 6  & 0 & 0 \\
Quest 2   & 9  & 1 & 0 \\
Quest Pro  & 9  & 1 & 0 \\
Quest 3 & 9  & 1 & 0 \\
Pico Neo 3   & 27 & 8 & 0 \\
Pico 4   & 27 & 7 & 0 \\
\hline
\end{tabular}
\caption{Security flags in the latest version of VR devices.}
\label{tab:manifest_flags}
\vspace{-12pt}
\end{table}

\subsubsection{Security flags in manifest files.}
\autoref{tab:manifest_flags} displays the current security flags present in the manifest files of the newest VR device models. 
Meta Quest devices have a low number of apps with the \cc{use\_cleartext\_traffic} flag set, reflecting a stricter stance against allowing unencrypted traffic.
Conversely, Pico devices have a much higher number of apps with \cc{use\_cleartext\_traffic} enabled, implying a more lenient approach to cleartext traffic that suggests potential security risks. 
For \cc{allow\_backup}, the original Quest device does not permit backups; however, Quest 2, 3, and Pro models provide for limited backup capabilities. In contrast, Pico devices have a much larger number of preinstalled apps allowing backup, indicating more liberal backup policies, which could pose privacy risks. Regarding the \cc{debuggable} flag, these are universally disabled across all devices, ensuring that debugging is turned off in production builds. \looseness=-1

\subsubsection{Permission analysis.}

We present the permission statistics in terms of different protection levels of the latest firmware versions in~\autoref{tab:permission}.
For the latest version, for Quest and Pico devices, the mean number of \cc{dangerous}, and \cc{normal} permissions requested per app was similar.
However,  Quest devices request a significantly larger number of \cc{others} permissions (2-3x) compared to Pico devices.  
This is due to the addition of a large number of custom permissions (\eg, 331 in Quest 3) to the Quest device family; 
Pico does not add these many custom permissions (\eg, 63 in Pico 4).
Moreover, the VR devices have introduced $17$ new \cc{dangerous} permissions, including \cc{FACE\_TRACKING}, \cc{BODY\_TRACKING}, and \cc{EYE\_TRACKING},
which can be requested by userspace applications. 

\begin{figure}
    \centering
    \includegraphics[width=0.8\linewidth]{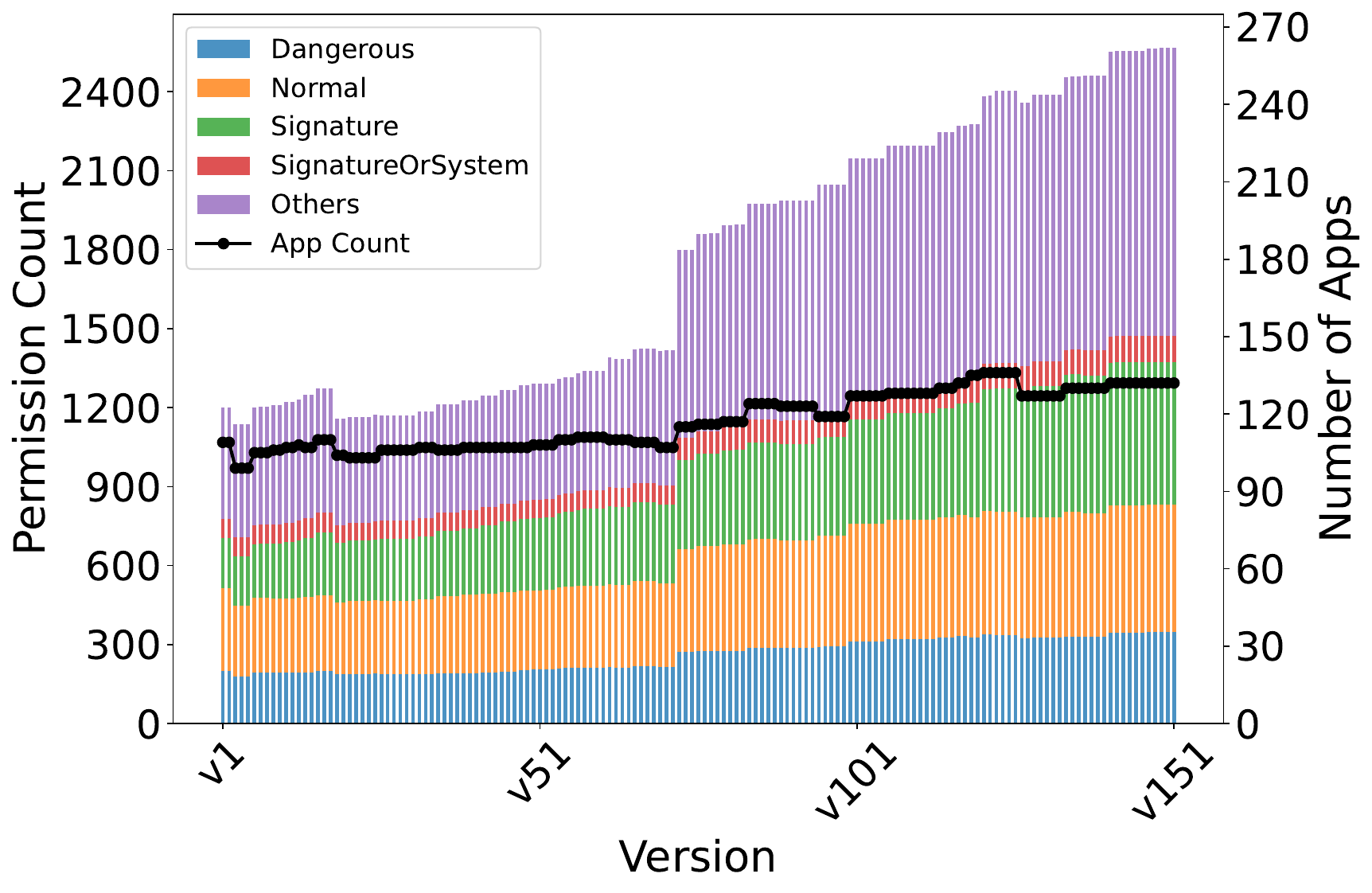}
    \vspace{-10pt}
    \caption{Permission     Changes of Quest 2}
    \label{fig:Quest2permissions}
    \vspace{-10pt}
\end{figure}

\bheading{Permission changes.}
For Pico devices, we did not observe many permission changes in their preinstalled apps, so we focus on Quest devices in the remainder of this section.
The longitudinal permission changes for Quest 2 are shown in~\autoref{fig:Quest2permissions} (similar figures for other devices are presented in \autoref{sec:app:rq2}).
We also plot the number of apps in the same figure.
We find that the number of \cc{dangerous} permissions have remained relatively stable across firmware versions, while
\cc{normal}, \cc{signature}, and \cc{SignatureOrSystem} permissions show steady growth over time.
\cc{others} permissions experienced a significant rise, escalating from approximately 300 to over 1000. This suggests a substantial number of custom permissions were introduced.
The most widely adopted permission longitudinally is \cc{READ\_PRIVILEGED\_PHONE\_STATE}. This permission is used to obtain the device identifiers. Other commonly requested permissions included \cc{ALWAYS\_CAPTURE\_MIC\_AUDIO\_INPUT}, 
\cc{RECORD\_AUDIO},  and \cc{START\_ACTIVITIES\_FROM\_BACKGROUND}. 

\begin{table}[t]
\footnotesize
\begin{tabular}{rrrrrr}
\toprule
Device    & Dangerous & Normal & Signature & SigOrSys& Others \\\midrule
Quest     & 2.02      & 2.93   & 2.77      & 0.7                 & 4.87   \\
Quest 2   & 2.63      & 3.67   & 4.09      & 0.75                & 8.30   \\
Quest Pro & 2.62      & 3.68   & 4.08      & 0.74                & 8.56   \\
Quest 3   & 2.64      & 3.64   & 4.04      & 0.74                & 8.17   \\
Pico Neo 3    & 1.95      & 2.92   & 1.33      & 0.62                & 2.86   \\
Pico 4    & 1.92      & 2.80   & 1.29      & 0.61                & 2.87  \\\bottomrule
\end{tabular}
\caption{Mean of permission counts of the latest firmware; SigOrSys: SignatureOrSystem; Others: all other permissions.}
\label{tab:permission}
\end{table}

\begin{figure}
    \centering
    \includegraphics[width=0.75\linewidth]{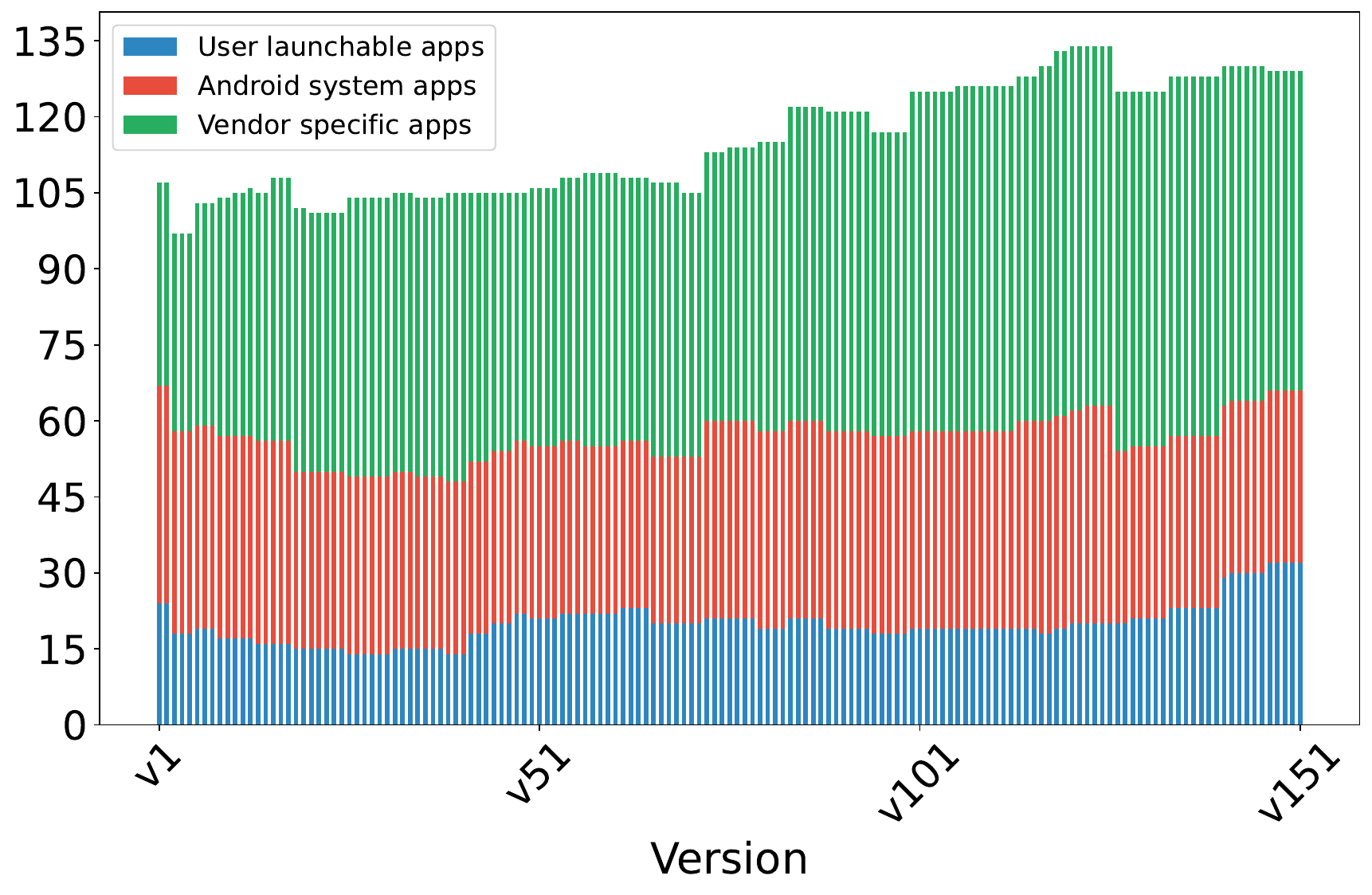}
    \vspace{-10pt}
    \caption{App Categorization of Quest2}
    \label{fig:types-Q2}
    \vspace{-10pt}
\end{figure}

\begin{figure}[t]
    \centering
    \includegraphics[width=0.75\linewidth]{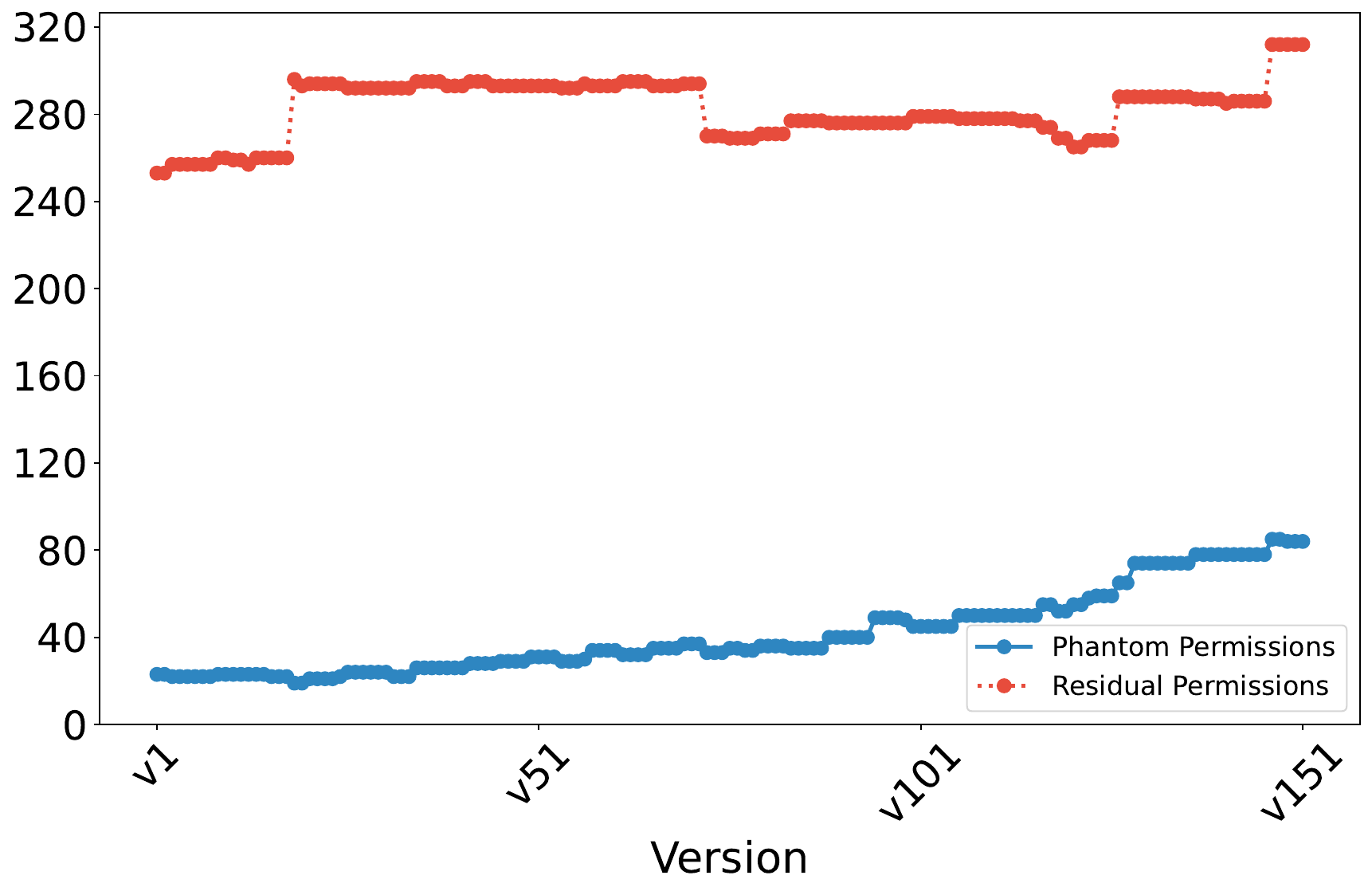}
    \vspace{-10pt}
    \caption{\textcolor{blue}{Phantom} and \textcolor{red}{Residual} Permissions of Quest 2}
    \label{fig:DNU}
\end{figure}

\begin{figure*}[htb]
    \centering
    \includegraphics[width=.9\linewidth]{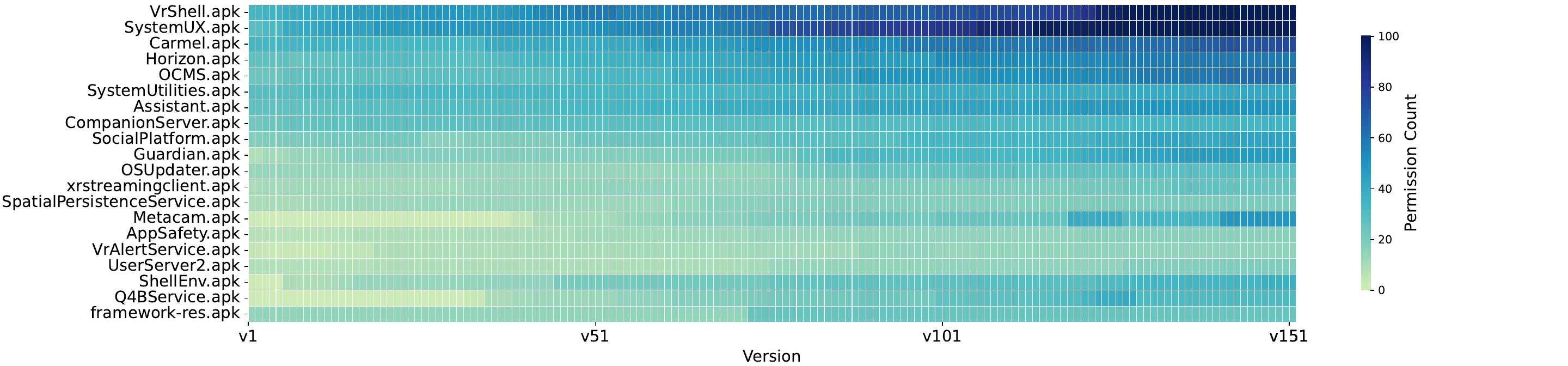}
    \vspace{-5pt}
    \caption{Top-20 Vendor apps with highest permission numbers in Quest 2.}
    \label{fig:Quest2Permission}
\end{figure*}

\bheading{Permission categorization based on app types.}
The app categorization of Quest 2 is shown 
in~\autoref{fig:types-Q2}.
Android system apps showed no significant increase in permission usage except during Android OS upgrades, while User launchable and Vendor specific apps showed significant increases across versions.
Vendor apps represent more than 50\% of the apps.
In~\autoref{fig:Quest2Permission},
we present top-20 Vendor apps with highest permission numbers in Quest 2 (similar figures for other devices are presented in~\autoref{sec:app:rq3}), which are selected from a total of 64 apps that are present across all firmware versions of Quest 2.
Several apps consistently show very high permission counts across all firmware versions,
such as SystemUX.apk, OCMS.apk and VrShell.apk. 

\bheading{Change of permission protection levels.}
In early versions of the Oculus Quest, we found that critical Android permissions such as \cc{REQUEST\_INSTALL\_PACKAGES} and \cc{SET\_TIME\_ZONE} were incorrectly marked as normal. Permissions with a normal protection level can be requested by any app without user approval, potentially allowing malicious apps to disrupt system behavior. With permissions like \cc{SET\_TIME\_ZONE}, third-party apps can disrupt logs, and with \cc{REQUEST\_INSTALL\_PACKAGES} permission, third-party apps can get the privilege to install other apps. In later versions, the permission protection level was changed to signature. Additionally, we identified that a custom permission \cc{RECORD\_AUDIO\_BACKGROUND} was initially assigned a dangerous protection level in Quest and Quest 2. Dangerous permissions can be granted to third-party apps with user consent, creating a privacy risk, especially for a permission that allows background audio recording. In later versions (Quest 3 and Quest Pro), this permission was completely removed, probably due to its sensitive nature.

Another concerning trend was the change in protection levels for permissions like \cc{FACE\_EYE\_INTERNAL\_API} and \cc{ORTHOFIT\_DATA}. These were originally marked as \cc{preinstalled} in Quest and Quest 2 — meaning only system apps could request them — but were later changed to \cc{dangerous}.
This shift allows third-party apps to request these sensitive permissions, increasing the potential for misuse. 

\bheading{Permission inconsistencies.}
We present the number of phantom and residual permission counts for preinstalled apps in Quest 2 over time in ~\autoref{fig:DNU}.
We can see that the phantom permissions increased over time (from 20 to over 80),
but residual permissions fluctuated (between 250 to 300).
An instance of residual permission is \cc{PERFORM\_SIM\_ACTIVATION}; although VR devices lack support for SIM cards, this permission is still specified at the framework level across all Quest device families.
A phantom permission example is \cc{CAR\_CONTROL\_AUDIO\_VOLUME}, a system-level permission for adjusting the car's audio volume when connected via Android Auto. This permission was applied in the Quest 2 within the \cc{VRSystemUI.apk} app without being declared,  and it was later removed through updates.
Another interesting phantom permission case is \cc{com.oculus.permission.HAND\_TRACKING}; this permission was used by preinstalled apps even before it was declared in the framework-level apk files, which may be the cause of some functionality bugs~\cite{reddit2023}.
Later, the permission was declared in the framework-level apk files. 

\begin{mdframed}[
  roundcorner=5pt,   linewidth=0.8pt,   linecolor=black,   backgroundcolor=gray!10,
  innertopmargin=4pt,innerbottommargin=4pt,
  innerleftmargin=4pt,
  innerrightmargin=4pt
]
\textbf{Summary.} 
VR devices are prone to vulnerabilities such as accidental data backups and exposure to MITM attacks due to vulnerable security flag configurations. Problems like permission inconsistencies, residual permissions, and phantom permissions highlight the security concerns in permission management.
\end{mdframed}
 
\section{RQ4: SEPolicy}
\label{s:rq4}

\subsection{Motivation}
SEPolicy is a set of rules used by SELinux to enforce Mandatory Access Control (MAC) within a system. It specifies how processes (subjects) can interact with other processes and system resources (objects). Vendors often customize the default Android SEPolicy to suit their devices. However, incorrect or overly permissive modifications can introduce security vulnerabilities. By default, SELinux adopts a {\it deny} approach, meaning any action not explicitly permitted by a rule is denied.
Analyzing SEPolicy for VR devices is crucial because these devices often include customized system components, sensors, and privileged services that aren't present in standard Android environments. These additions may require vendor-specific policy changes, which, if misconfigured, could unintentionally weaken security boundaries—especially given the sensitive data and capabilities (e.g., camera, motion tracking) involved in VR platforms.

\subsection{Methodology}
In the firmware images, the SEPolicy files (\cc{.cil} files) are present in \cc{/etc/selinux} in \cc{system.img} and \cc{/etc/selinux} in \cc{vendor.img}. 
After extracting the \cc{.cil} files,
we have processed them to extract the \cc{allow}  (explicitly permit specific actions) and \cc{never-allow} rules (prevent certain permissions from being granted).
As early Quest 1 firmware (running Android 7) only contains SEPolicy binaries instead of \cc{.cil} files, we exclude them from our study; we only include 24 Quest 1 firmware ranging from April 2021 to December 2023 (running Android 10). 

\subsection{Longitudinal Analysis}
\begin{figure}
    \centering
    \includegraphics[width=0.9\linewidth]{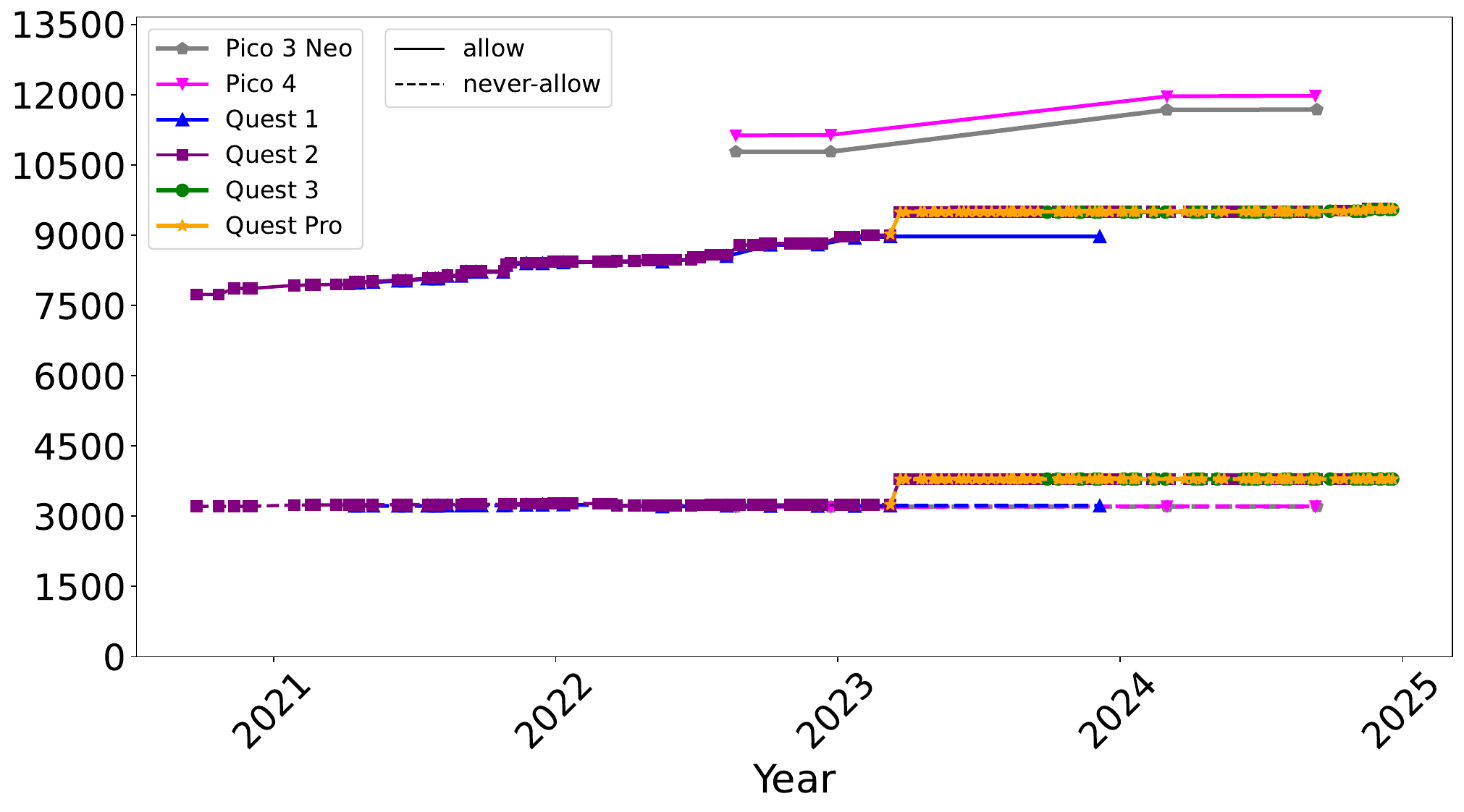}
    \vspace{-10pt}
    \caption{SEPolicy: Allow and Never-allow Rules}
    \label{fig:selinux}
\end{figure}

\autoref{fig:selinux} shows the evolution of SEPolicy rules across different VR devices.
Longitudinally,
the number of  \cc{never-allow} rules remains stable for all devices.
The only increase appeared in 2023 for Quest 2, where
it upgraded from Android 10 to 12.
For \cc{allow} rules,
the numbers grow gradually for Quest and Pico devices.
There are many more  \cc{allow} rules than  \cc{never-allow} rules for all devices.

\bheading{Isolated and untrusted app domains.}
Isolated and untrusted app domains in SEPolicy are intended to regulate access by third-party applications. 
Through our analysis of \cc{allow} and \cc{never-allow} rules targeting these domains, we observed notable differences between Quest and Pico devices. 
Specifically, all devices within the Quest family retained the default Android SEPolicy configurations, showing no customizations.
In contrast, Pico devices exhibit significant changes to these policies. Pico has explicitly removed certain \cc{never-allow} rules associated with isolated and untrusted app domains.
The removal of such rules weakens the SEPolicy's rigor on Pico devices. For example, in Pico devices, the never-allow rules corresponding to \cc{net\_dns\_prop}, \cc{radio\_cdma\_ecm\_prop} were removed.
Moreover, Pico allows untrusted applications to access vendor-specific property files such as \cc{vendor\_android\_pvr\_prop},  which is a property file introduced by Pico.
Additionally, another property, \cc{hwservicemanager\_prop}, typically restricted by Android, is accessible to untrusted apps on Pico. These property files usually contain unique device identifiers, and their exposure poses privacy risks such as device fingerprinting~\cite{dong2025insecurity, meng2023post}.

\begin{mdframed}[
  roundcorner=5pt,   linewidth=0.8pt,   linecolor=black,   backgroundcolor=gray!10,
  innertopmargin=4pt,innerbottommargin=4pt,
  innerleftmargin=4pt,
  innerrightmargin=4pt
]
\textbf{Summary.} 
VR devices often customize SEPolicies. We found that some \cc{never-allow} rules have been removed in VR devices, which may lead to security issues. 
Additionally, applications from untrusted domains can access property files, potentially leading to privacy issues like device fingerprinting.
\end{mdframed}

 \section{Discussion}
\label{sec:discuss}

Our longitudinal study of VR firmware reveals several recurring and systemic security weaknesses across kernel configuration, binary hardening, application privilege enforcement, and SELinux policy design.
While some protections are partially adopted, the overall security posture of VR firmware lags behind modern Android standards.
Below, we offer concrete recommendations for VR device vendors and discuss the limitations of our study.

\subsection{Impacts of Our Findings}
\rev{
Lack of proper kernel mitigations (RQ1) can allow privilege escalation attacks that tamper with raw sensor data (e.g., eye-tracking) or camera feeds. Similarly, weak hardening (RQ2) and mismanaged permissions (RQ3) can let third-party or preinstalled apps exploit privileged system services, posing risks in sensitive VR applications such as medical or military training.
In addition, incorrect or overly permissive modifications of SEPolicy (RQ4)
can introduce security vulnerabilities to the VR platforms, leading to privacy violations (\eg, leakage of sensitive motion data) or privilege escalations.}

\rev{We found that VR-specific permissions such as \cc{ORTHOFIT\_DATA} and \cc{FACE\_EYE\_INTERNAL\_API}, used for tracking, were changed from \cc{preinstalled} to \cc{dangerous}, exposing sensitive data to third-party apps and reflecting emerging attack surfaces not present in traditional Android systems. In addition, SEPolicy customizations in VR firmware can enable untrusted apps to access vendor-defined properties related to motion tracking. We also observed that binaries introduced in recent VR firmware (e.g., \cc{libsensorutils}, \cc{pvr\_compute}) often lacked standard hardening protections, suggesting a trade-off between real-time responsiveness and security, that is specific to VR. }

\subsection{Responses from Vendors}
\rev{Following responsible disclosure practices, we reported our findings to Meta and Pico, who have acknowledged our findings. While both vendors requested proof-of-concept exploits to validate the security implications, the closed nature of these VR platforms precludes dynamic analysis capabilities, as detailed in~\autoref{sec:discuss:limit}. Consequently, our empirical study identifies potential security and privacy vulnerabilities through static analysis, rather than demonstrating active exploits.
These findings underscore systemic weaknesses in VR firmware security that merit further investigation.
}

\bheading{Pico's response.}
\rev{Through iterative communications with Pico's development team, we received comprehensive technical responses regarding our identified security concerns.}
\rev{Pico, like other OEM vendors, inherits kernel source code directly from SoC manufacturers such as Qualcomm. The kernel version is intrinsically coupled with the underlying SoC hardware, making kernel upgrades infeasible without corresponding hardware changes. This architectural constraint explains why Pico Neo 3 and Pico 4, both utilizing Qualcomm XR2 SoC, maintain the same 4.19 kernel version. The development team has committed to maintaining security through continuous integration of upstream vulnerability patches.}

\rev{
The vendor noted that binary hardening mechanisms (Stack Canaries, CFI, Fortify, NX, RELRO) alone do not determine program security, as runtime context and program functionality must also be considered. Their security policy mandates hardening only for high-risk binaries that expose external services or have known vulnerabilities. The vendor cited technical challenges in implementing universal hardening across their large codebase, including legacy code compatibility issues and high engineering costs associated with modifying existing binaries.}

\rev{
Regarding \cc{use\_cleartext\_traffic} flag, Pico explained that despite the flag being enabled, their proprietary network library enforces HTTPS, ensuring encrypted network traffic. 
They also noted that \cc{allow\_backup} is an intended Android OS feature rather than a vulnerability, though we analyzed it following prior work~\cite{guo2024empirical}.
}
\rev{Pico provides SDKs that enable third-party developers to access device features and enable SEPolicy rules. Their security team is actively reviewing and refining SEPolicy configurations to eliminate unnecessary rule modifications in future releases.
}

\bheading{Meta's response.}
\rev{
Meta provided targeted feedback on specific findings. They clarified that applications with \cc{use\_cleartext\_traffic} enabled are inherited from AOSP defaults and largely inactive in HorizonOS. For sensitive permissions like \cc{ORTHOFIT\_DATA} and \cc{FACE\_EYE\_INTERNAL\_API}, Meta implemented system-level access controls restricting usage to authorized system applications. Their permission deployment strategy involves initial testing with first-party applications before broader third-party availability, which may explain the observed permission level transitions.
}

\subsection{Recommendations for  Stakeholders}

\bheading{For vendors.}
\rev{
A consistent theme across all VR firmware we analyzed is the partial and inconsistent adoption of core Android security mechanisms.
This is mainly because VR firmware development currently lacks a standardized, enforceable baseline like Android’s Compatibility Definition Document (CDD).
As a result, vendors vary significantly in their implementation of core security practices.
We recommend the VR vendors adopt a formal compliance framework that specifies minimum requirements, such as kernel configurations, compiler flags, SELinux policies, and permission enforcement rules, modeled on the Android CDD but specific to VR.
Strict SEPolicy isolation should be mandated for new VR system services, with least-privilege boundaries clearly enforced.
}

\rev{
The limited use of advanced mitigations in VR may reflect perceived tradeoffs between performance and security, especially in latency-sensitive VR environments.
Prior works~\cite{Sidhpurwala_RELRO, zhang2013practical, maar2024defects} have demonstrated that hardening techniques, such as CFI, full RELRO, or updated kernel configurations, can introduce overhead that may affect rendering performance or real-time sensor processing.
However, our findings suggest that vendors are making these tradeoffs without fully accounting for the security implications.
We recommend that in cases where vendors disable mitigations due to performance constraints, these decisions should be transparently documented and justified.
Alternative defenses, such as stronger sandboxing, isolation, or monitoring, should be employed to compensate.
Performance-aware variants of mitigations (e.g., lightweight CFI) should also be explored.
For instance, Site Isolation in Chrome~\cite{reis2019site} demonstrates how performance and security can be jointly optimized in large-scale platforms.}

\rev{
Our analysis of preinstalled apps and SELinux policies reveals that configurations often remove \cc{neverallow} rules, introduce unused domains (e.g., telephony services on non-cellular VR devices), or fail to isolate new capabilities.
Preinstalled apps frequently request excessive permissions, including access to sensitive data (e.g., \cc{FACE\_TRACKING} and \cc{EYE\_TRACKING}), with protection levels changing across updates. These patterns suggest inadequate review and enforcement.
We recommend that vendors regularly audit app permissions and SELinux policies, apply strict protection levels to custom permissions, and isolate all new services using fine-grained SELinux domains.
Legacy and unused policies should be aggressively pruned to reduce the attack surface and improve policy clarity.
}

\bheading{For developers.}
\rev{
VR applications are often built using complex game engines and development frameworks (e.g., Unity, Unreal), which offer rapid development workflows but may enable insecure defaults.
Our analysis shows that many apps depend on untrusted third-party components for core functionality such as motion tracking and input handling.
We recommend that developers disable insecure defaults (e.g., \cc{allowBackup=true}, overly broad intent filters) and avoid exposing sensitive components unnecessarily.
When using Unity or Unreal, build configurations should be explicitly hardened following platform guidance (e.g., through ProGuard, and code obfuscation).
In addition, developers should avoid incorporating third-party libraries for privacy-sensitive functionality (e.g., eye tracking, facial capture, and spatial tracking) unless they are thoroughly vetted. 
Developers should request only those permissions essential to app functionality and avoid over-scoping permission declarations, especially for biometric or sensor data.
}

\bheading{For manufacturers.}
\rev{
Many modern VR headsets are built on top of mobile SoCs that include hardware-backed security features (e.g., pointer authentication, TEEs).
However, we found that these capabilities are often not enabled or not fully integrated into production firmware.
We recommend that chipset manufacturers collaborate closely with headset manufacturers to ensure that hardware capabilities are fully leveraged and enabled by default in VR firmware builds.
This includes providing reference firmware configurations, performance benchmarks, and integration guidelines to reduce adoption barriers.
}

\subsection{Limitations}
\label{sec:discuss:limit}

\bheading{No source code available for VR devices.}
VR devices run on tailored Android systems, with modifications being much more extensive than those seen in standard Android smartphones.
Despite this, no VR device manufacturer has publicly released its source code. This makes it extremely challenging, if not impossible, to infer the underlying reasons for our findings observed in firmware. 
For instance, consider the situation involving residual permissions; had the source code been accessible to us, we could have conducted a more thorough analysis regarding the presence of these residual permissions and their potential implications.
This limitation also restricts our ability to study customizations in the framework layer.

\bheading{No dynamic analysis.}
Currently, gaining root access on commercial VR devices is not feasible (Except Oculus Go~\cite{meta_oculus_go_unlocking}).
As a result, we are unable to escalate privileges or gain system-level access required to perform dynamic analysis,
which is achievable in prior Android firmware papers~\cite{lee2021polyscope}.
In addition, major VR vendors do not i) provide official simulators, or ii) expose low-level APIs or debugging endpoints for fuzzing or instrumentation.
This limitation restricts our ability to monitor runtime behaviors, detect dynamic privilege use, or uncover real-time data flows and interprocess communication. Consequently, our analysis is constrained to a static approach.
\looseness=-1

\bheading{Limited VR firmware datasets.}
In our study, we focused primarily on the VR firmware of the Meta Quest Series and Pico. Although there are additional VR devices available, such as the HTC Vive and Varjo, we were unable to locate any available resources regarding their firmware. Due to the lack of root access, extracting firmware images directly from the devices is highly challenging. A possible solution involves capturing the over-the-air updates provided to these VR devices, which include the updated firmware images. We leave a thorough exploration of such methods as our future work.

\bheading{Generalizability.}
\rev{
While our research questions and findings are designed for Android-based VR platforms, and can extend to future Android-XR devices, they may not apply to headsets running fundamentally different OS (e.g., Apple’s visionOS).}

 \section{Related Work}
\label{sec:related}

\bheading{Android firmware analysis.}
Existing works have performed analysis on Android firmware to study security issues.
Several works carry out measurement studies on hundreds of Android firmware from different smartphone vendors~\cite{gamba2020analysis,hou2022large,hou2023can}.
Researchers have also proposed different firmware analysis tools
to study different aspects such as pre-installed apps~\cite{sutter2023firmwaredroid,elsabagh2020firmscope},
security policies~\cite{hernandez2020bigmac},
AT commands~\cite{tian2018attention},
init routines~\cite{ji2021definit}
and unix domain sockets~\cite{elgharabawy2022sausage}.

\bheading{Android customization.}
Another important aspect of Android is the customization introduced by device vendors,
as demonstrated by existing works~\cite{wu2013impact,zhou2014peril,aafer2016harvesting,yu2021sepal,possemato2021trust,el2021dissecting,liu2022customized,jin2023dependency}.
Third-party libraries are an important component of the Android ecosystem.
Many existing works aim to identify third-party libraries and their vulnerabilities in Android apps~\cite{zhan2020automated,backes2016reliable,derr2017keep,li2017libd,zhang2019libid,wang2023union,xie2023precise}.
In our work,
we focus on analyzing firmware of VR systems.

\bheading{VR application analysis.}
Other researchers have assessed VR apps or SDKs to explore privacy concerns. In OVRSeen~\cite{trimananda_ovrseen_2022}, the authors conducted an investigation into privacy breaches in Oculus applications by analyzing network traffic. Subsequently, a tool named VPVet~\cite{zhan2024vpvet} was developed to evaluate the privacy policies of VR applications. 
A recent study~\cite{guo2024empirical} has examined third-party VR apps. Different from these works, our research targets a longitudinal analysis of VR firmware.

\bheading{VR attacks.}
Previous studies have revealed that privacy leakages in VR environments can be exploited for user deanonymization attacks. Most of these studies utilize built-in sensors (e.g., accelerometer and gyroscope) in VR headsets and analyze user motion data~\cite{miller2022combining,liebers2021understanding,olade2020biomove,pfeuffer2019behavioural,tricomi2022you,kumar2022passwalk,zhang2023s,nair2022exploring,slocum2023going,aziz2025exploring,jarin2023behavr}, whereas a few of them employ specialized external devices to explore other user data \cite{ishaque2020physiological,vsalkevicius2019anxiety}, e.g., electroencephalogram. 
The attacker may also utilize side channels to obtain sensitive information. 
Secrets such as keystrokes in VR
can also be inferred
from motion data~\cite{zhang2023s,luo2022holologger,slocum2023going,meteriz2022keylogging} and other side-channel information,
such as videos of user movements~\cite{gopal_hidden_2023,nguyen_penetration_2024,yang_can_2023},
acoustic signals~\cite{luo_eavesdropping_2024}, network traffic~\cite{su_remote_2024} and
WiFi signals~\cite{al2021vr}.
Recent works~\cite{nair_deep_2023, nair_going_2023, li2021kalvarepsilonido} have proposed injecting noise into motion data or identifiable anthropometrics to make inference attacks harder,
which have predominantly used differential privacy~\cite{dwork_differential_2006,dwork_differential_2008}.
However, these works did not study the firmware of VR systems.

\section{Conclusion}
In this paper, we conduct the {\it first} in-depth security analysis of more than 300 VR firmware over time from two major vendors, Quest and Pico.
Our study uncovers critical security issues within VR firmware, including absent kernel and binary protections, inconsistent permission settings, and improper removal of SELinux policy enforcements. This research sheds light on the largely uncharted area of VR firmware security, providing valuable insights that can guide the development of future VR systems.

\section*{Acknowledgement}
The authors from George Mason University (GMU) are supported in part by 1) a seed funding and GRA awards from the CAHMP (now CHAIS) Center at GMU, and 2) a seed funding from 4-VA, a collaborative partnership for advancing the Commonwealth of Virginia. 
The author from the University of California, Irvine (UCI) is supported by the UCI Academic Senate Council on Research, Computing, and Libraries (CORCL) Award.

{
\bibliographystyle{IEEEtranS}
\bibliography{paper}
}

\begin{figure*}[t]
  \centering
  \begin{minipage}[t]{0.32\textwidth}
    \centering
    \includegraphics[width=\linewidth]{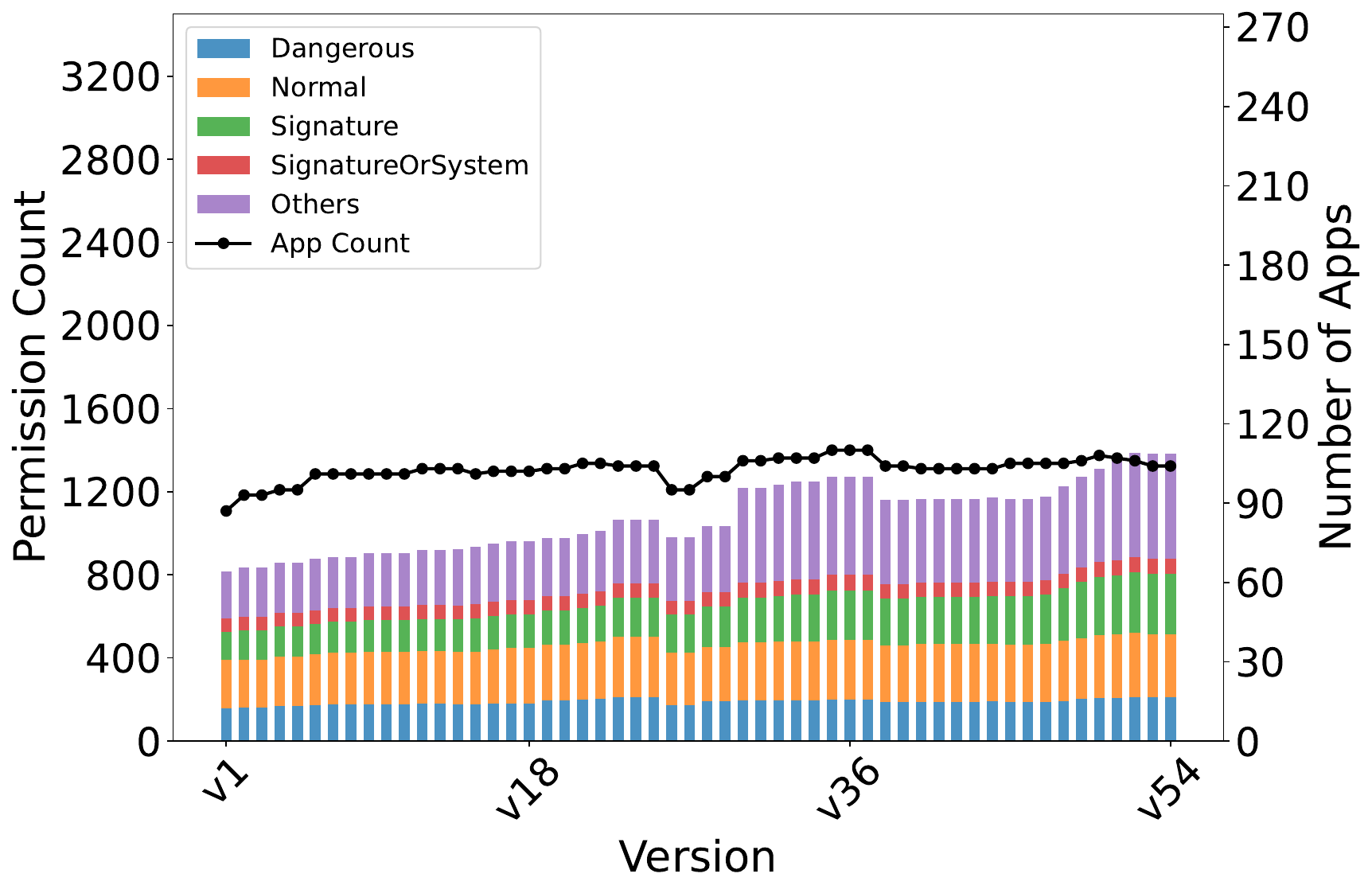}
    \vspace{-10pt}
    \caption{Quest 1 permissions}
    \label{fig:Quest1permissions}
  \end{minipage}\hfill
  \begin{minipage}[t]{0.32\textwidth}
    \centering
    \includegraphics[width=\linewidth]{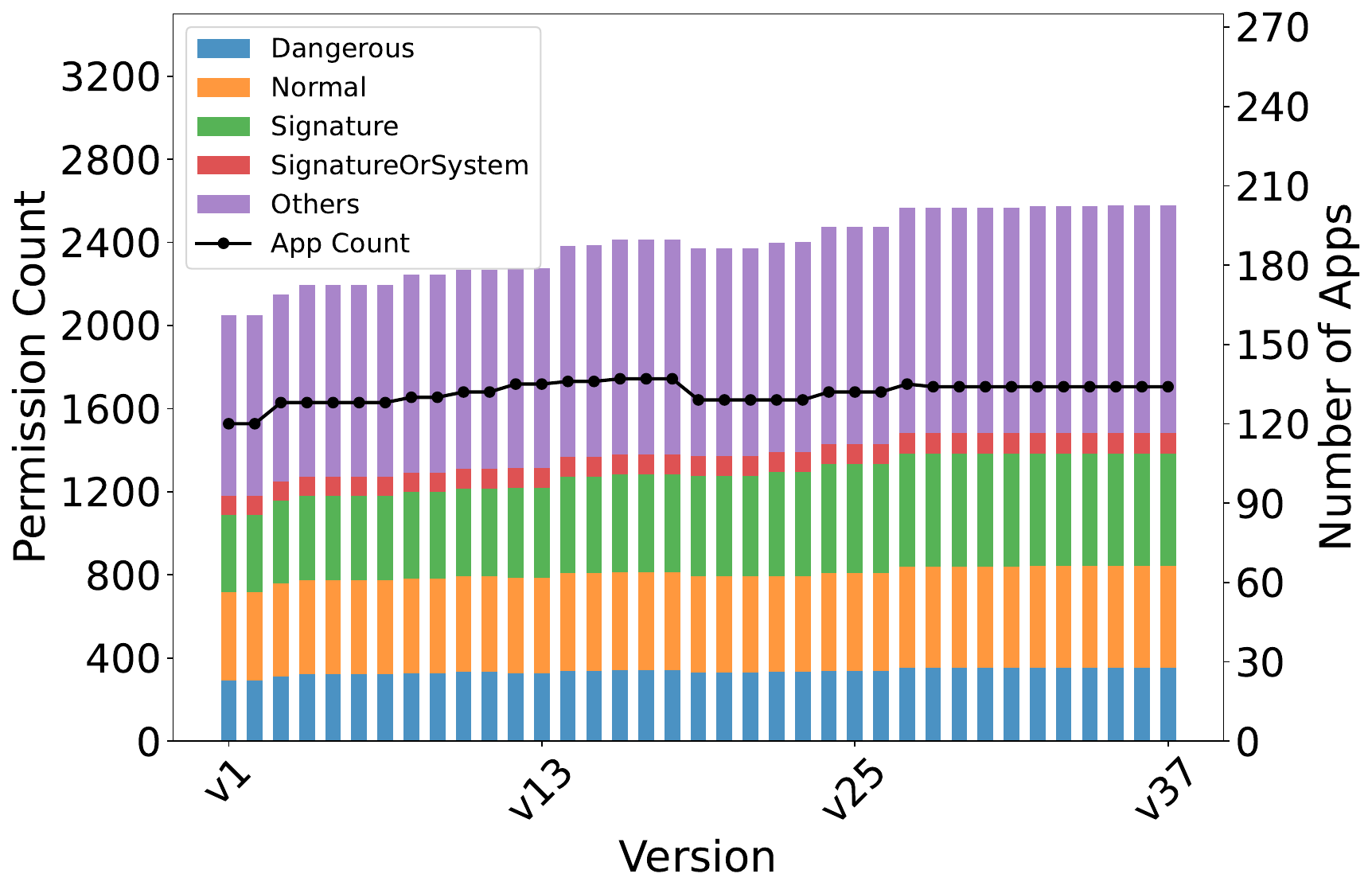}
     \vspace{-10pt}
    \caption{Quest 3 permissions}
    \label{fig:Quest3permissions}
  \end{minipage}\hfill
  \begin{minipage}[t]{0.32\textwidth}
    \centering
    \includegraphics[width=\linewidth]{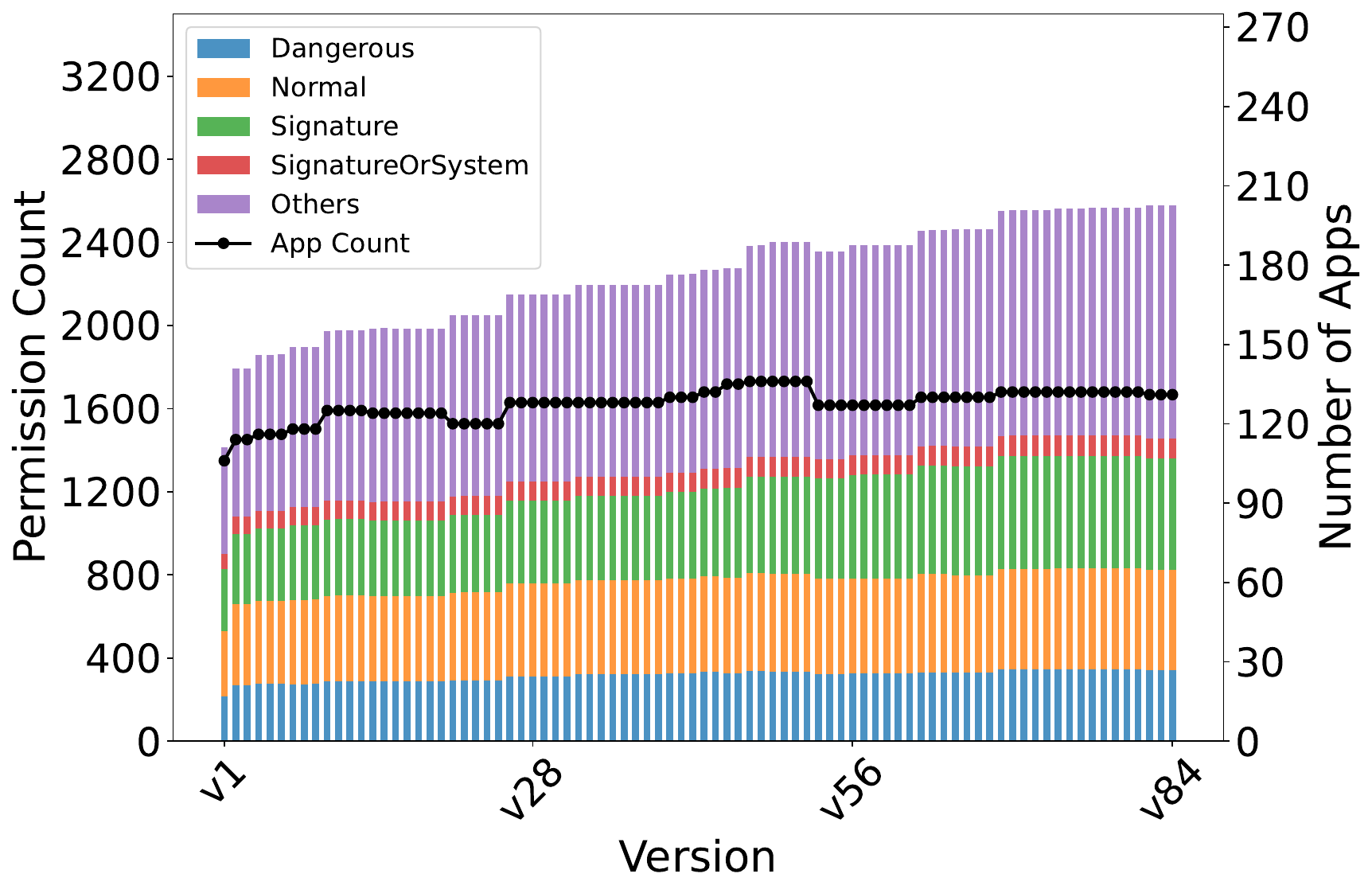}
     \vspace{-10pt}
    \caption{Quest Pro permissions}
    \label{fig:QuestPropermissions}
  \end{minipage}
\end{figure*}

\begin{figure*}[t]
  \centering
  \begin{minipage}[t]{0.32\textwidth}
    \centering
    \includegraphics[width=0.9\linewidth]{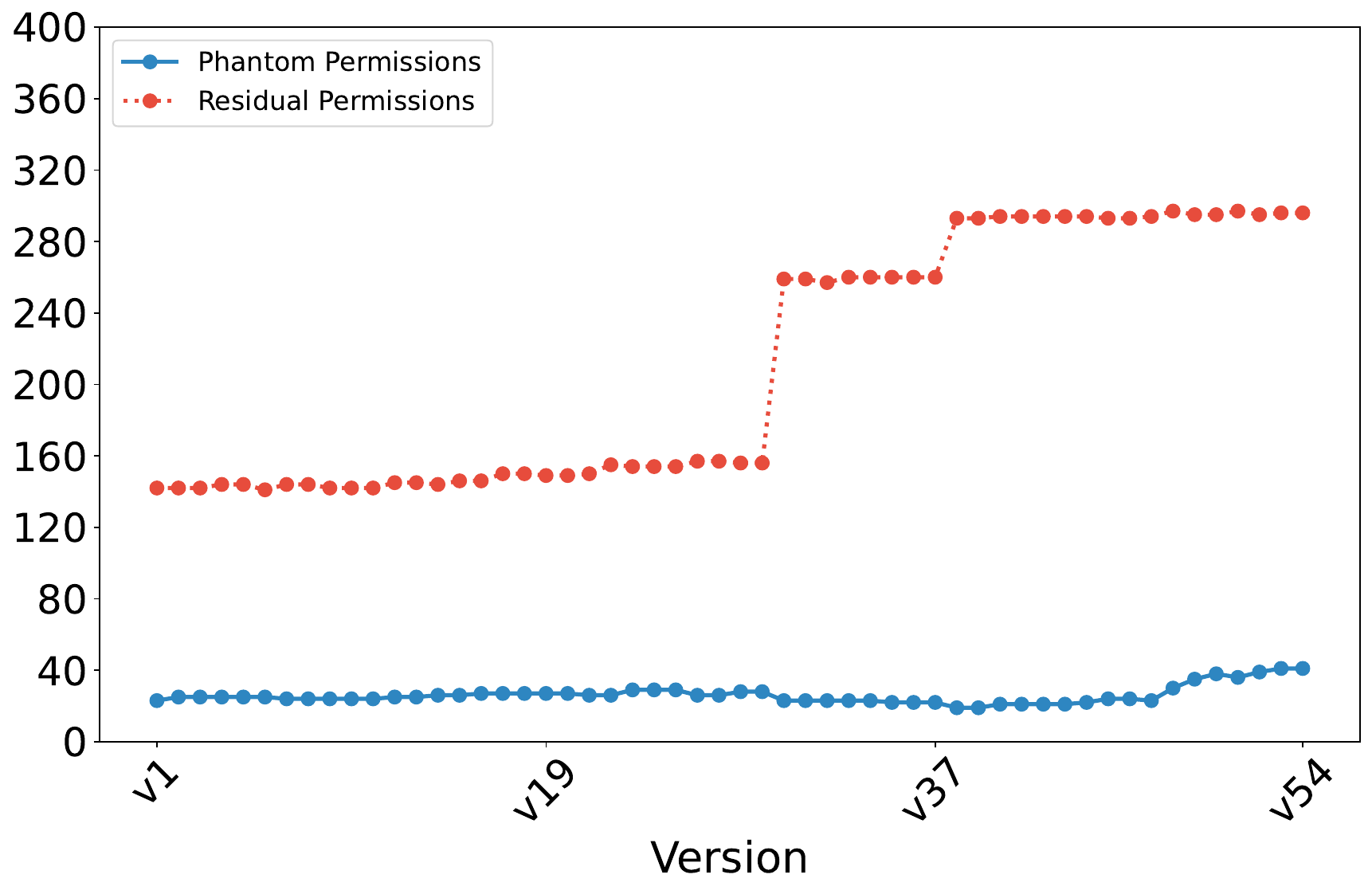}
    \caption{\textcolor{blue}{Phantom} and \textcolor{red}{Residual} Permissions of Quest 1}
    \label{fig:DNU-Q1}
  \end{minipage}\hfill
  \begin{minipage}[t]{0.32\textwidth}
    \centering
    \includegraphics[width=0.9\linewidth]{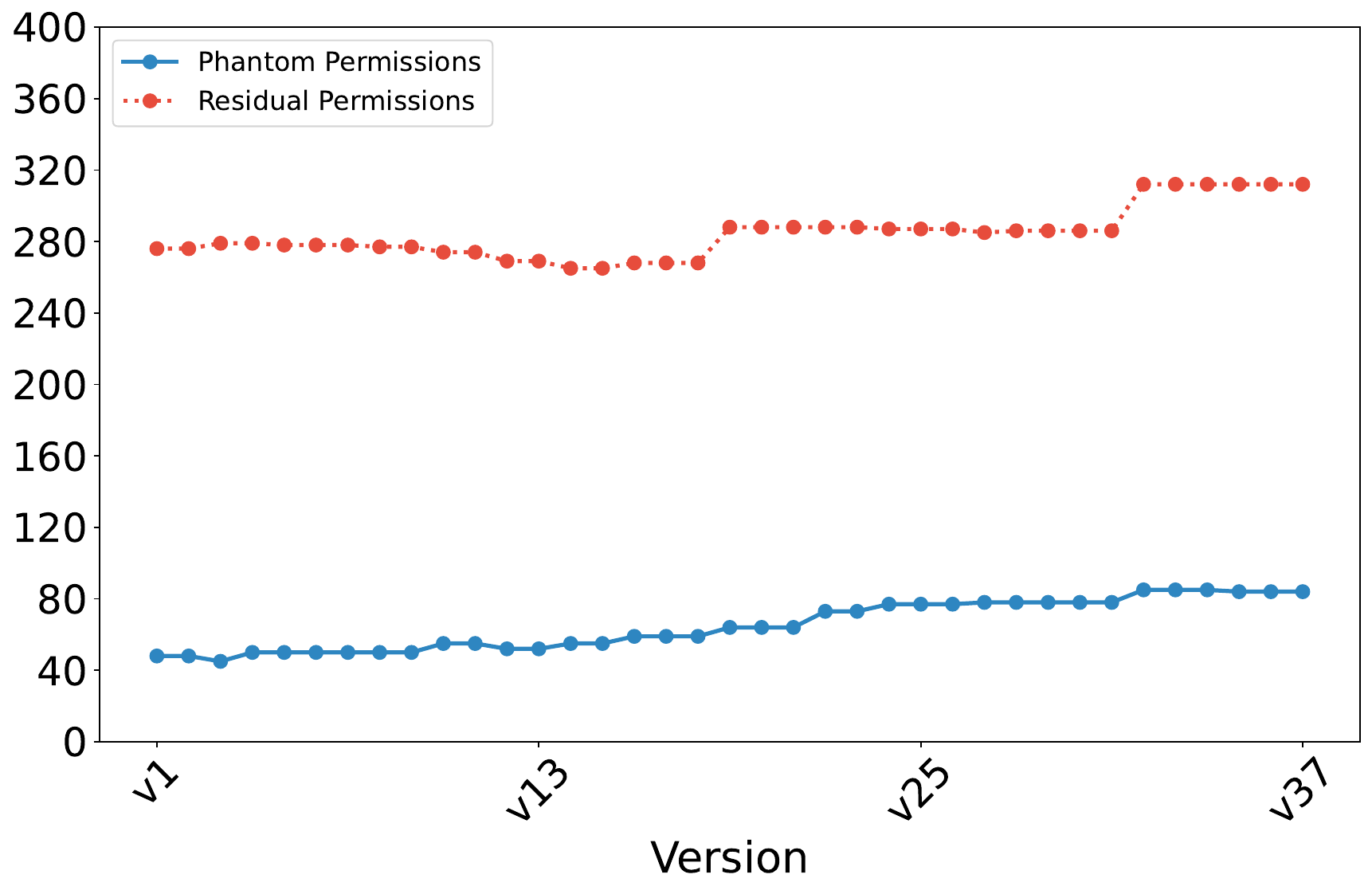}
    \caption{\textcolor{blue}{Phantom} and \textcolor{red}{Residual} Permissions of Quest 3}
    \label{fig:DNU-Q3}
  \end{minipage}\hfill
  \begin{minipage}[t]{0.32\textwidth}
    \centering
    \includegraphics[width=0.9\linewidth]{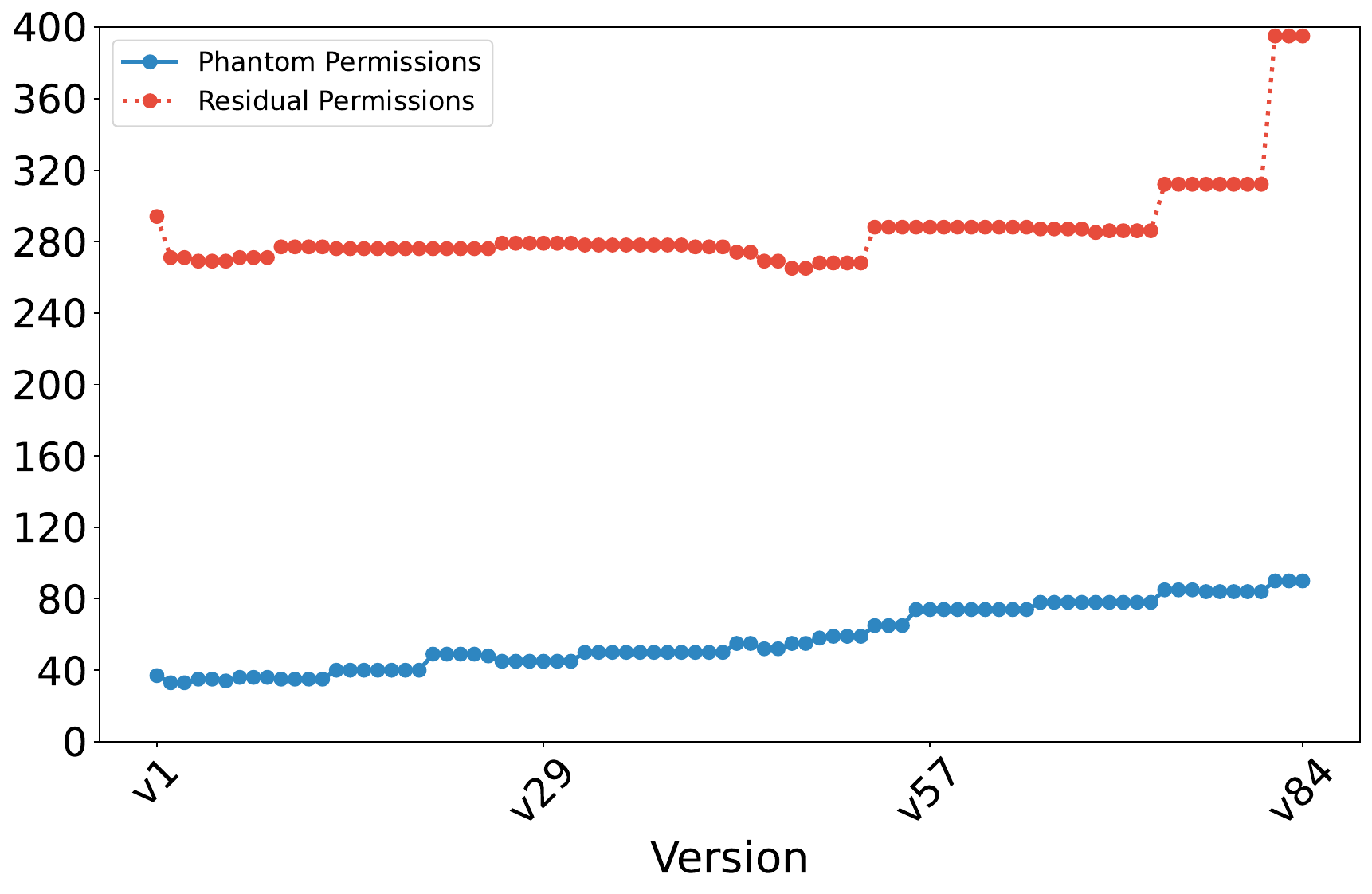}
    \caption{\textcolor{blue}{Phantom} and \textcolor{red}{Residual} Permissions of Quest Pro}
    \label{fig:DNU-Qpro}
  \end{minipage}
\end{figure*}

\newpage
\appendix

\section{RQ2: Latest Version Analysis}
\label{sec:app:rq2}

\begin{table}[htbp!]
\footnotesize
\begin{tabular}{cccccc}
\toprule
Device    & No Canaries & No CFI & No Fortify Source & No NX & No RELRO   \\\midrule
Quest     &   323    &2712& 1687&  7   &  7   \\
Quest2    &   451    &2757& 1893& 38   & 38    \\
Quest Pro &   443    &2750& 1907&  39  & 39     \\
Quest 3   &   436    &2909& 1950& 10   & 10      \\
Pico 3 Neo   &   242    &4161& 2417&   77  & 67      \\
Pico 4    &   219    &4207& 2071&   5  & 13         \\       \bottomrule
\end{tabular}
\caption{Binaries with no hardening in the latest version.}
\label{t:binary}
\vspace{-10pt}
\end{table}

\autoref{t:binary} summarizes the number of binaries that lack specific mitigations on the latest firmware version of each device. 
{\bf CFI} and {\bf Fortify Source} are the most frequently missing protections across all devices. This is especially concerning, as both mechanisms play critical roles in preventing control-flow hijacking and buffer overflows, respectively.
For instance, Pico 3 and Pico 4 exhibit high counts of binaries lacking CFI, with $4161$ and $4207$ binaries missing this protection, respectively. Similarly, they also show poor adoption of Fortify, with 2411 and 2071 binaries lacking it. 
{\bf RELRO}  and {\bf NX} are generally well-adopted; 
the number of binaries lacking these protections is consistently low across all devices, ranging from 7 to 13 for RELRO and 7 to 77 for NX. 
When comparing {\bf Canaries}, Pico devices show better deployment than the Quest devices. 
For example, Pico 3 and Pico 4 have 242 and 219 binaries lacking canary support, which is lower than all quest devices. 
This suggests a stronger emphasis on stack protection in the Pico firmware.

\section{RQ3: Additional Results}
\label{sec:app:rq3}

\begin{table*}[ht]
\centering
\footnotesize
\begin{tabular}{ll}\toprule
\textbf{Build version} & \texttt{ro.build.id} property for Quest devices; \texttt{ro.build.release} and \texttt{ro.system.build.version.sdk} in \texttt{/system/build.prop} for Pico devices \\ \hline
\textbf{Kernel binary location}   & \texttt{boot.img} \\ \hline
\textbf{Binaries location}        & \texttt{/system/bin}, \texttt{/system/apex}, \texttt{/system/lib}, \texttt{/system/lib64}, \texttt{/vendor/bin}, \texttt{/vendor/lib}, \texttt{/vendor/lib64}, \texttt{/odm/bin}, \texttt{/odm/lib}, \texttt{/odm/lib64} \\ \hline
\textbf{Apps location}            & \texttt{/system/priv-app/}, \texttt{/system/app/}, \texttt{/vendor/app}, \texttt{/odm/app} \\ \hline
\textbf{SEPolicy location}        & \texttt{/system/etc/selinux}; \texttt{/etc/selinux} \\
\bottomrule
\end{tabular}
\caption{System Paths and Properties}
\label{t:system_paths}
\vspace{-10pt}
\end{table*}

\begin{figure*}[t]
  \centering
  \begin{minipage}[t]{0.32\textwidth}
    \centering
    \includegraphics[width=0.9\linewidth]{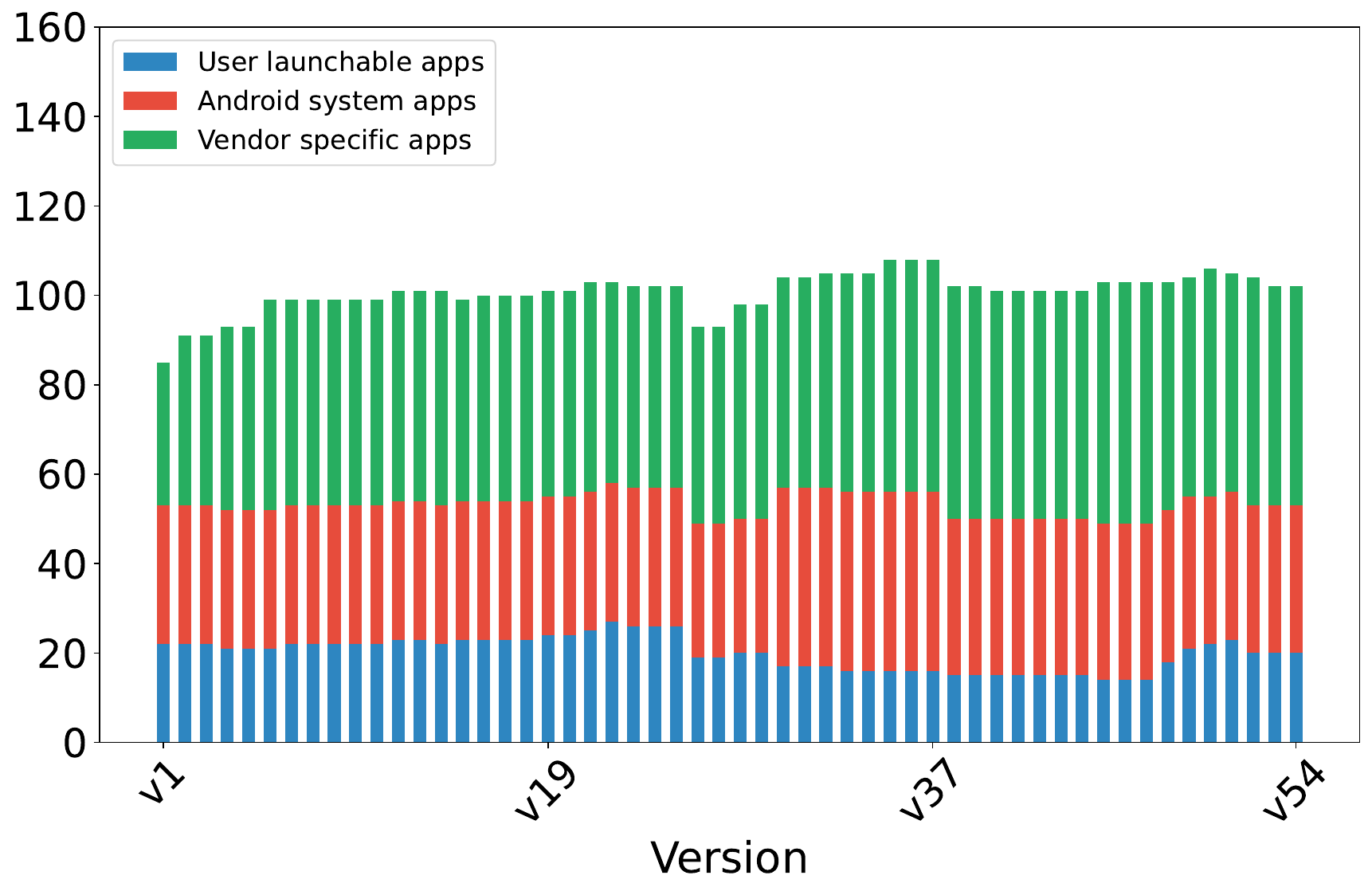}
    \vspace{-10pt}
    \caption{App Categorization of Quest 1}
    \label{fig:type-Q1}
    \vspace{-10pt}
  \end{minipage}\hfill
  \begin{minipage}[t]{0.32\textwidth}
    \centering
    \includegraphics[width=0.9\linewidth]{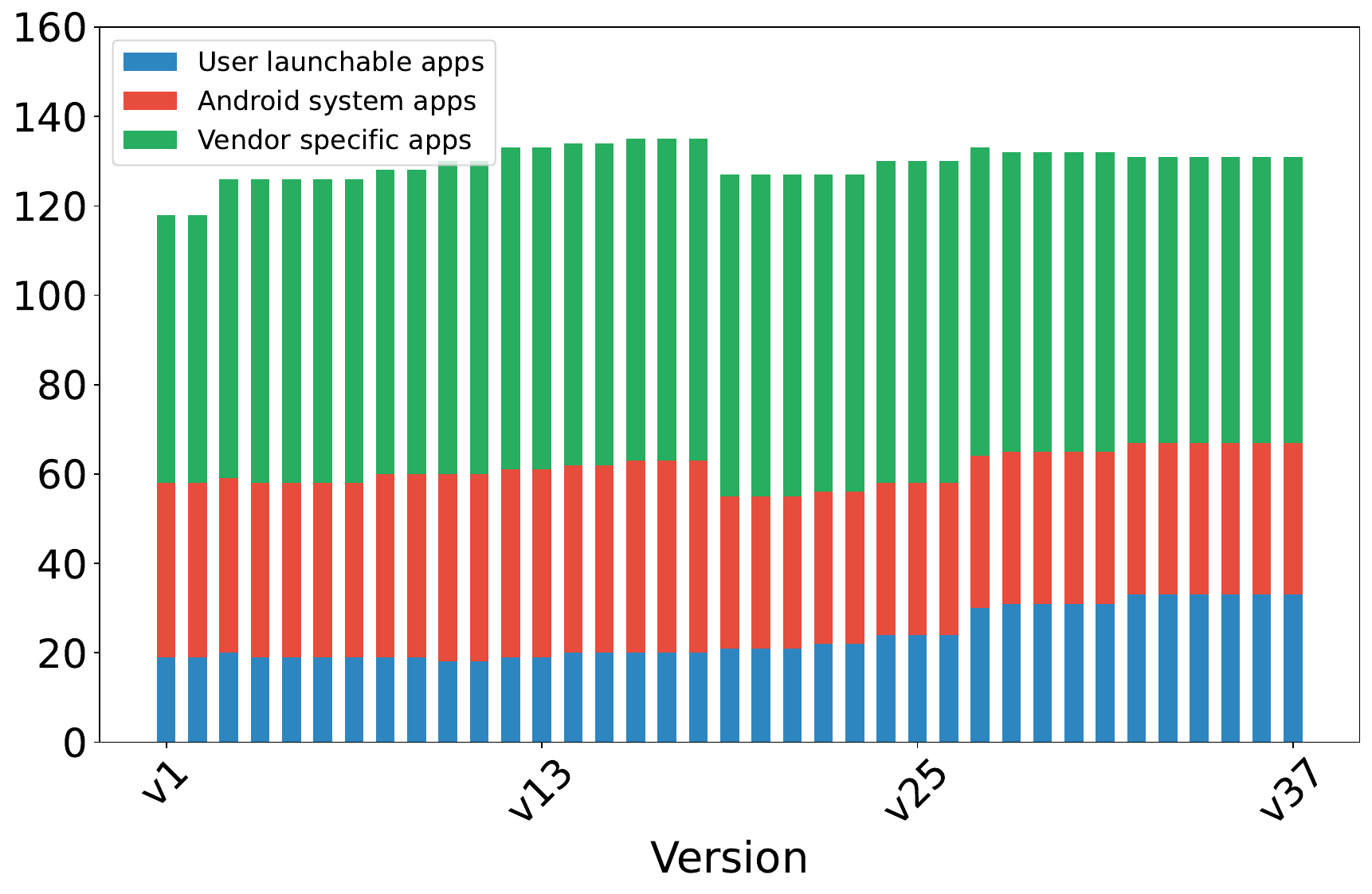}
    \vspace{-10pt}
    \caption{App Categorization of Quest 3}
    \label{fig:type-Q3}
    \vspace{-10pt}
  \end{minipage}\hfill
  \begin{minipage}[t]{0.32\textwidth}
    \centering
    \includegraphics[width=0.9\linewidth]{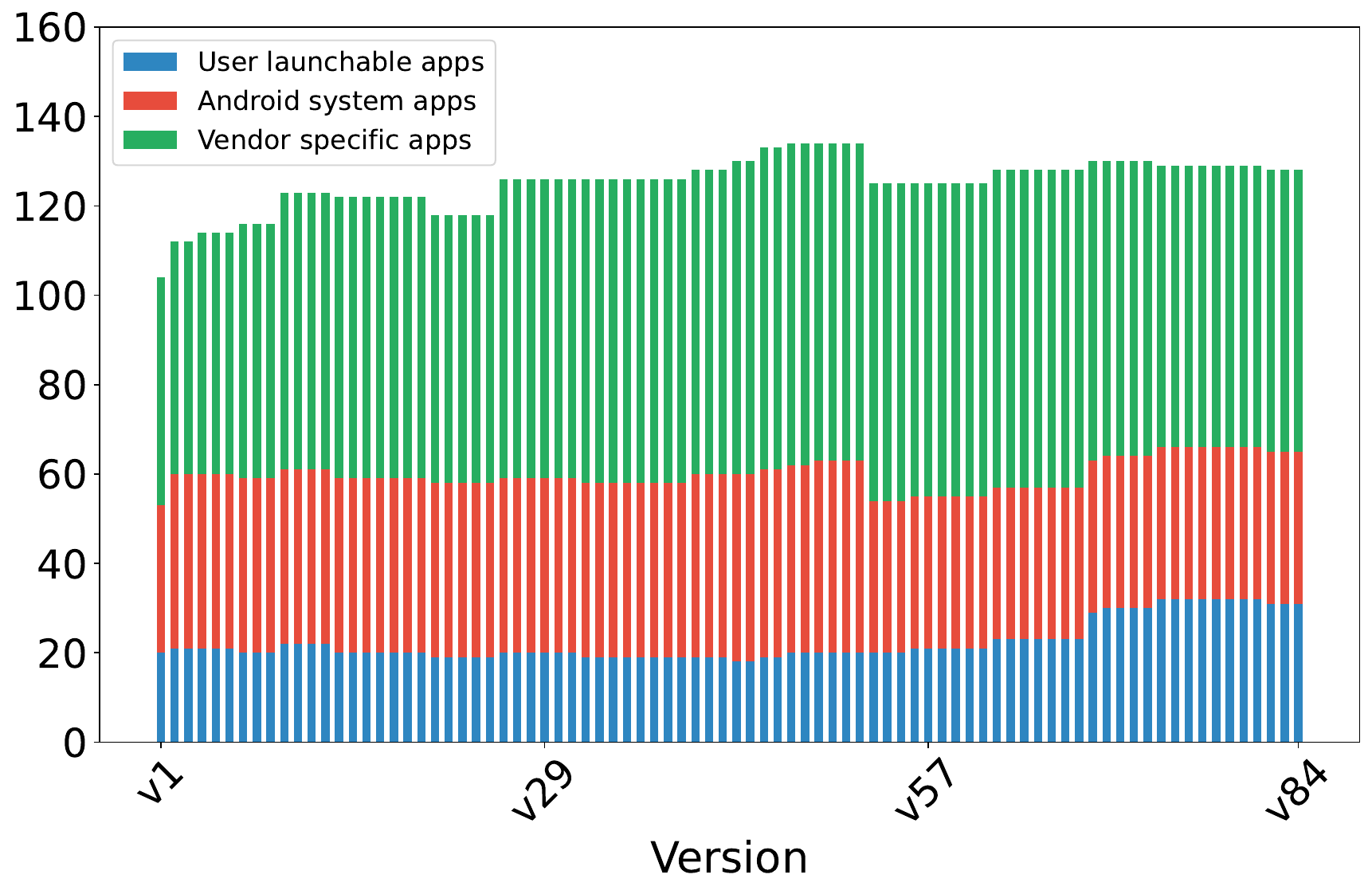}
    \vspace{-10pt}
    \caption{App Categorization of Quest Pro}
    \label{fig:type-Qpro}
    \vspace{-10pt}
  \end{minipage}
\end{figure*}

Here we present the RQ3 results on Quest, Quest 3, and Quest Pro.

\subsection{Longitudinal Analysis}

\bheading{Quest.}
~\autoref{fig:Quest1permissions} shows permission changes for Quest 1. Dangerous permissions remain relatively stable throughout all versions. Normal permissions and Signature permissions show modest growth over time. The ``Others'' category  demonstrates the most significant increase, suggesting substantial addition of custom permissions across firmware updates. 

\bheading{Quest 3.}
~\autoref{fig:Quest3permissions}  depicts permission trends for Quest 3. This device shows higher baseline permission counts overall compared to Quest 1. Dangerous, Normal, and Signature permissions maintain relatively stable values across versions. The ``Others'' category shows consistent growth, increasing from around 1800 to approximately 2600. 

\bheading{Quest Pro.}
~\autoref{fig:QuestPropermissions} presents Quest Pro permissions. Dangerous permissions remain steady while Normal and Signature permissions show slight increases. The ``Others'' category experiences the most dramatic growth, nearly doubling from about 700 to 1400 permissions.

\begin{figure}[t]
    \centering
    \includegraphics[width=.99\columnwidth]{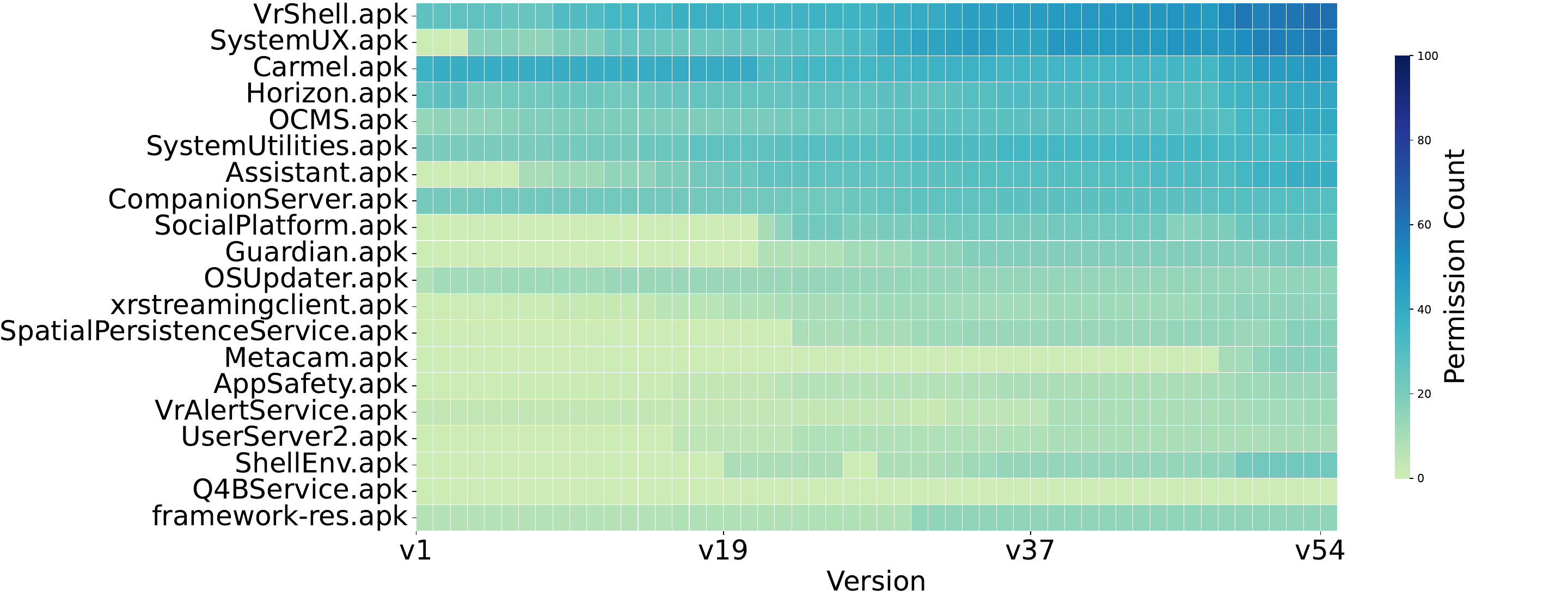}
    \vspace{-10pt}
    \caption{Top-20 vendor apps with highest permission numbers in Quest 1.}
    \label{fig:heatmap-Q1}
    \vspace{-10pt}
\end{figure}

\begin{figure}[t]
    \centering
    \includegraphics[width=.99\columnwidth]{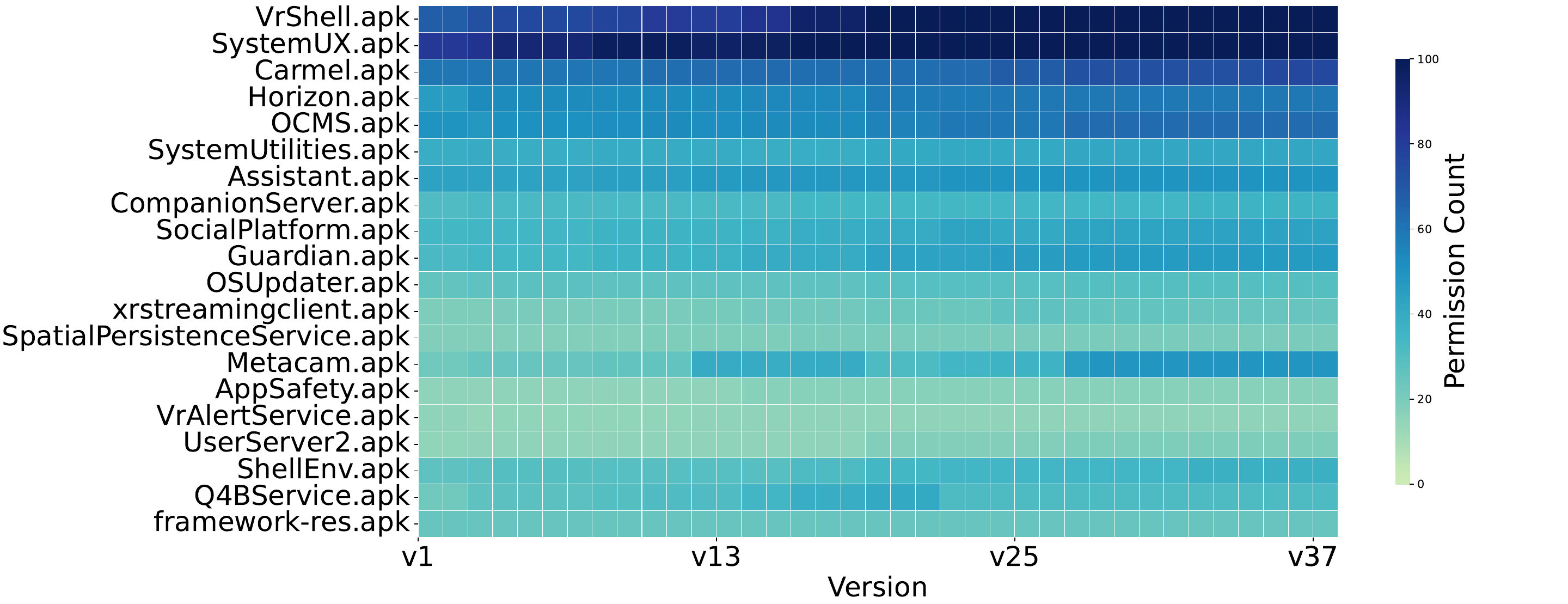}
    \vspace{-10pt}
    \caption{Top-20 vendor apps with highest permission numbers in Quest 3.}
    \label{fig:heatmap-Q3}
    \vspace{-10pt}
\end{figure}

\begin{figure}[t]
    \centering
    \includegraphics[width=.99\columnwidth]{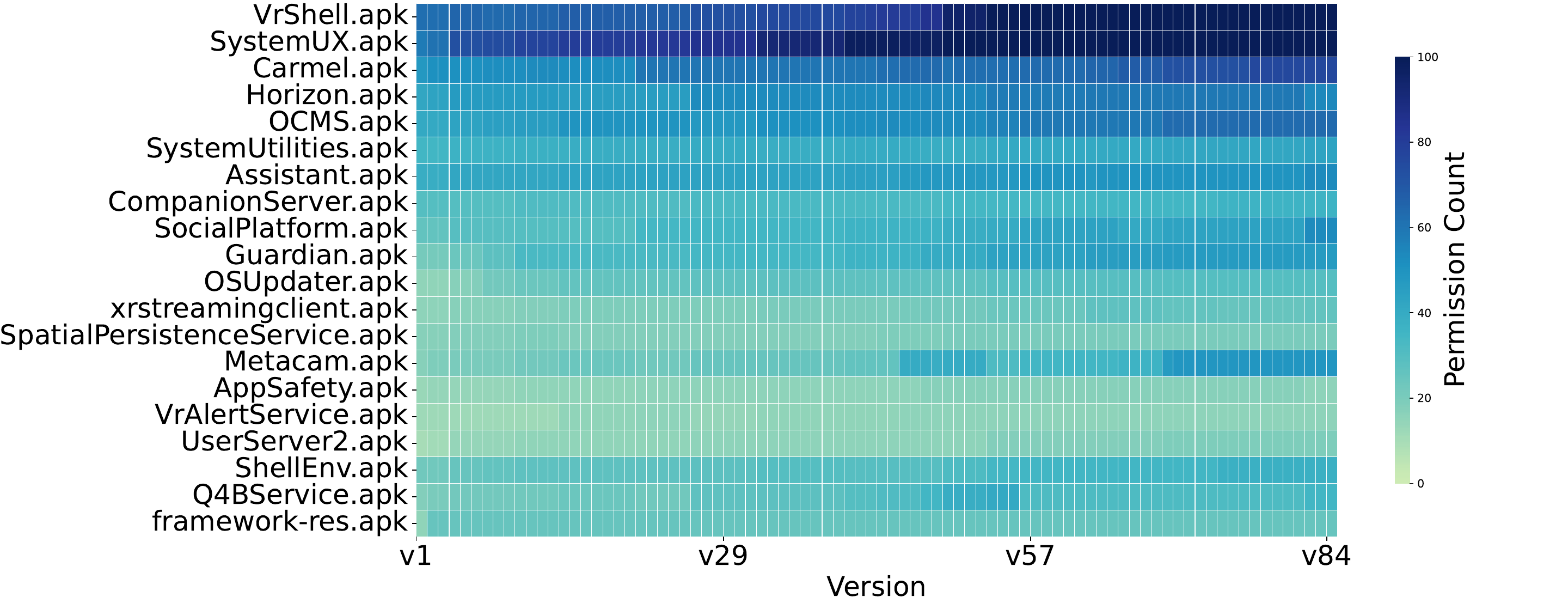}
    \vspace{-10pt}
    \caption{Top-20 vendor apps with highest permission numbers in Quest Pro.}
    \label{fig:heatmap-Qpro}
\end{figure}

\subsection{Permission Categorization}

\bheading{Quest.}
In~\autoref{fig:type-Q1},
we present the app categorization result of Quest 1.
User launchable and Android system apps remain relatively stable.
Vendor specific apps comprise the largest portion with minor fluctuations.
In~\autoref{fig:heatmap-Q1},  we present permission counts for top-20 APKs across Quest 1 firmware. Several apps consistently show very high permission counts, such as VrShell.apk which maintains the highest counts across all firmware versions. SystemUX.apk and SystemUtilities.apk also exhibit significant permission usage, while most applications display gradual permission increases rather than sudden changes. 

\bheading{Quest 3.}
The app categorization of Quest 3 is presented in~\autoref{fig:type-Q3}.
User launchable apps (with UI) show a notable increase in later firmware versions, indicating a greater focus on user interface applications over time.
Android system apps remain relatively stable throughout.
Vendor specific apps  make up the largest portion across all firmware versions, but their proportion decreases slightly in later versions.
In~\autoref{fig:heatmap-Q3}, we examine top-20 Vendor specific apps in Quest 3, revealing that VrShell.apk and SystemUX.apk begin with already elevated permission counts and maintain consistently high levels throughout this version range. 

\bheading{Quest Pro.}
The app categorization of Quest 3 is presented in~\autoref{fig:type-Qpro}. 
User launchable apps increase gradually;
Android system  apps remain relatively consistent;
Vendor apps show some growth.
In ~\autoref{fig:heatmap-Qpro}, we present permission distribution of top-20 vendor apps of Quest Pro. VrShell.apk and SystemUX.apk demonstrate exceptionally high permission counts (near 100) from the earliest versions, with Carmel.apk showing moderately high permissions that increase in later versions. 

\subsection{Permission Inconsistencies}

\bheading{Quest.}
\autoref{fig:DNU-Q1} shows phantom and residual permissions across Quest 1 firmware versions. Phantom permissions remain relatively stable at first, and later they increase a bit. The residual permissions display a dramatic jump at v29 because of Android update, followed by another smaller increase later on.

\bheading{Quest 3.}
\autoref{fig:DNU-Q3} depicts permission trends for Quest 3. Phantom permissions show a gradual upward trajectory. Residual permissions fluctuates a bit, then increase.

\bheading{Quest Pro.}
~\autoref{fig:DNU-Qpro} presents Quest Pro's permission evolution. Phantom permissions demonstrate a contiguous growth pattern, beginning near 40 and incrementally increasing to nearly 90. Residual permissions exhibit more dramatic changes; it starts with very stable trend, then shows a substantial increase.
 \end{document}